\documentclass[aps,prb,twocolumn,10pt,superscriptaddress,showpacs]{revtex4-2}
\usepackage{amsmath,amssymb,graphics,epsfig,epstopdf,color,verbatim,ulem,braket,tabularx,multirow,array,diagbox,colortbl,hhline}
\usepackage[colorlinks,linkcolor=blue,citecolor=blue,urlcolor=blue]{hyperref}
\usepackage[italicdiff]{physics}
\usepackage{bm}
\usepackage{blindtext}
\newcommand{\eyps}{\boldsymbol{s}}
\newcommand{\gpps}{\Omega}

\begin{document}

\title{Computation of thermal entropy for the doped Hubbard model}

\author{Yu-Feng Song}
\affiliation{Institute of Modern Physics, Northwest University, Xi'an 710127, China}
\affiliation{Hefei National Laboratory for Physical Sciences at Microscale and Department of Modern Physics, University of Science and Technology of China, Hefei, Anhui 230026, China}

\author{Youjin Deng}
\email{yjdeng@ustc.edu.cn}
\affiliation{Hefei National Laboratory for Physical Sciences at Microscale and Department of Modern Physics, University of Science and Technology of China, Hefei, Anhui 230026, China}
\affiliation{Department of Modern Physics, University of Science and Technology of China, Hefei, Anhui 230026, China}
\affiliation{Hefei National Laboratory, Hefei 230088, China}

\author{Yuan-Yao He}
\email{heyuanyao@nwu.edu.cn}
\affiliation{Institute of Modern Physics, Northwest University, Xi'an 710127, China}
\affiliation{Shaanxi Key Laboratory for Theoretical Physics Frontiers, Xi'an 710127, China}
\affiliation{Fundamental Discipline Research Center for  Quantum Science and Technology of Shaanxi Province, Xi'an 710127, China}
\affiliation{Hefei National Laboratory, Hefei 230088, China}

\begin{abstract}
We develop a highly efficient framework for computing the thermal entropy in the doped Fermi-Hubbard model within the grand-canonical ensemble. The framework comprises four calculation schemes that express the entropy as path integrals in the parameter space of temperature, interaction strength, and chemical potential. The integrands involve only fundamental observables, including the total energy, fermion density, and double occupancy, which are readily accessible in a wide range of theoretical and numerical methods. We further derive useful Maxwell relations connecting the entropy to other quantities and present practical formulas for directly evaluating the grand potential. As an application, we compute the entropy of the doped Hubbard model in two and three dimensions, using the numerically unbiased auxiliary-field quantum Monte Carlo method. The test results show excellent agreement across the different schemes and quantitatively verify the Maxwell relations, confirming the reliability of the framework. In two dimensions, we further benchmark our entropy results in physically relevant parameter regimes against diagrammatic Monte Carlo calculations and observe excellent quantitative consistency between the two approaches. By providing an efficient and broadly applicable route for entropy evaluation, our work facilitates the thermodynamic characterization of complex correlated states in the doped Hubbard model.
\end{abstract}
\maketitle

\section{Introduction}
\label{Sec:Intro}

In correlated quantum systems, thermodynamic quantities encode signatures of microscopic correlations and collective phenomena, and are widely used as theoretical and experimental probes of quantum criticality and emergent order~\cite{Sachdev2011,Georges1996,Masatoshi1998,Lohneysen2007,Bloch2008,Navon2010}. Among them, thermal entropy is particularly important, as it quantifies the number of accessible microstates and captures the interplay between quantum correlations and thermal fluctuations~\cite{Garst2005,Rost2009,Rost2011,Grube2017,Ryll2014}. It exhibits characteristic behaviors in a variety of quantum states and phases, including the Fermi liquid, {Mott} insulator, quantum spin liquids, antiferromagnets, and superconductors~\cite{Gorelik2012,Song2025L,*Song2025B,Lu2026,Balents2010,Wessel2010,Dagotto1994,Bulut2002}. The entropy landscape provides essential insights into cryogen-free cooling protocols, such as adiabatic demagnetization refrigeration near quantum criticality in magnetic materials~\cite{Bernd2011,Xiang2024,Shu2026,Lin2026} and interaction-induced adiabatic cooling (also known as Pomeranchuk cooling) in Fermi-Hubbard systems~\cite{Werner2005,Dar2007,Paiva2011,Leo2011,GangLi2014,Song2025CPL}. In ultracold atomic experiments, thermal entropy is a more fundamental metric than temperature for quantifying how cold the system is~\cite{Bloch2012,Ho2009,Joerdens2008,Schneider2008,Duarte2015,Tarruell2010,Daniel2013,Hart2015,Daniel2015,Shao2024}, since temperature is a derived quantity and is typically not directly measurable. More broadly, the entropy lies at the interface of quantum many-body physics and information-theoretic concepts~\cite{Islam2015,Cocchi2016,Cocchi2017,Walsh2019L,*Walsh2019B,Walsh2020,Walsh2021}.

In the Fermi-Hubbard model~\cite{Hubbard1963,Kanamori1963,Gutzwiller1963}, the paradigmatic model for strongly correlated electrons~\cite{Arovas2022,Qin2022}, thermal entropy plays a significant role in understanding the thermodynamic properties and associated underlying physics of the system~\cite{Song2025L,*Song2025B,Song2025CPL,Lu2026,Khatami2009,Khatami2011,Sordi2011,Lenihan2021}. At half filling, entropy is one of the key observables used to characterize the metal-to-{Mott} insulator crossover in both two- and three-dimensional ({2D} and 3D) models~\cite{Song2025L,*Song2025B,Lu2026}. Its nonmonotonic behavior with interaction strength controls the temperature dependence of double occupancy via a Maxwell relation, and essentially leads to the Pomeranchuk cooling effect~\cite{Song2025CPL,Lu2026}. Away from half filling, the early studies showed that~\cite{Khatami2009,Khatami2011,Sordi2011}, at sufficiently strong interactions, the variation of entropy with doping displays the signature of a possible quantum critical point in the {2D} Hubbard model. This doping dependence was re-examined more recently in Ref.~\cite{Lenihan2021}, where an additional crossover from a non-Fermi-liquid (or pseudogap) state to a Fermi-liquid state was identified. These previous results demonstrate that the evolution of thermal entropy in the multidimensional parameter space is intimately connected to the full phase diagram of the Hubbard model, which remains a central focus of condensed matter theory and computational quantum many-body physics. Hence, the efficient computation of thermal entropy is crucial for the study of the Hubbard model.

In our previous work~\cite{Song2025L,*Song2025B}, we established the formalism for computing thermal entropy for the half-filled standard Hubbard model. We refined the conventional calculation scheme based on the integration over temperature~\cite{Dar2007,Ibarra2020,Sushchyev2022,Paiva2010}, and developed a novel formalism via an integration over interaction strength. The latter can substantially reduce the computational cost when evaluating the entropy as a function of interaction strength at fixed temperature. With doping, the additional parameter of chemical potential $\mu$, which controls the fermion filling $n$ of the model, is introduced (within the grand canonical ensemble), and accordingly renders the computation of entropy more complicated.

Although numerous existing studies~\cite{Leo2011,Imriska2016,LeBlanc2013,GangLi2014,Lenihan2021,Paiva2011,Fuchs2011,Rampon2025,Khatami2009,Khatami2011,Sordi2011,Ehsan2011,Khatami2012,Paiva2010,Walsh2019L,*Walsh2019B,Cocchi2017,Walsh2020,Walsh2021} have involved entropy evaluation in the doped Hubbard model, the techniques applied all have certain limitations. The {\it first} approach~\cite{Imriska2016,Leo2011,LeBlanc2013,GangLi2014,Paiva2010, Paiva2011,Fuchs2011,Rampon2025,Khatami2009,Khatami2011,Sordi2011} is to simply employ the scheme of integration over temperature $T$ (or equivalently integration over inverse temperature $\beta=1/k_BT$), as widely used in the half-filled case. Nevertheless, such calculations must deal with the high-temperature limit ($T=\infty$), which usually involves a high-$T$ cutoff, and may also suffer from the coarse resolution in $\beta$ at low temperatures. Both issues can compromise the numerical accuracy. Moreover, the entropy formulas in this approach only work for calculations with fixed fermion fillings, and need to be slightly modified for the fixed-$\mu$ situation (see Secs.~\ref{Sec:FixUn} and~\ref{Sec:FixUmu}). The {\it second} route~\cite{Ehsan2011,Khatami2012,Lenihan2021} directly computes the entropy from the grand potential via their connection formula. However, this approach is only practical in specific many-body numerical methods, including numerical linked cluster expansions (NLCE) and diagrammatic Monte Carlo (DiagMC), which can directly access the grand potential. The {\it third} method obtains thermal entropy in two steps~\cite{Walsh2019B,Cocchi2017,Walsh2020,Walsh2021}. It first numerically calculates the pressure via an integration of fermion density $n(\mu,T)$ over $\mu$, from the empty state [$\mu=+\infty$ based on the model Hamiltonian in Eq.~(\ref{eq:Hamiltonian})] to the specific $\mu$ value at a desired filling. Then the entropy is the temperature derivative of the pressure. This procedure requires performing wide $\mu$-sweep calculations (even with a practical $\mu$ cutoff), and involves successive numerical integration and differentiation, which may lead to instability issues. Thus overcoming the above limitations and developing simple yet generally applicable schemes for computing thermal entropy in the doped Hubbard model remains highly desirable.

In this work, we address the efficient evaluation of thermal entropy in the doped Hubbard model and develop a unified computational framework for this purpose. It incorporates four complementary calculation schemes that all formulate the entropy as simple path integrals over temperature, interaction strength, and chemical potential. The integrands involve very fundamental observables, including the total energy, fermion density, and double occupancy. We derive these entropy formulas independently from the definition of the grand potential and the finite-temperature Hellmann-Feynman theorem. We also obtain useful Maxwell relations that connect the entropy to other thermal quantities. The framework should be broadly applicable to various theoretical and numerical methods, and we demonstrate its applications here using auxiliary-field quantum Monte Carlo (AFQMC) simulations for both the {2D} and 3D doped Hubbard model. We find excellent agreement among the results obtained from different schemes within the framework, as well as with the physically relevant results reported in previous DiagMC calculations.

The rest of this paper is organized as follows. In Sec.~\ref{Sec:MMO}, we first describe the doped Hubbard model and briefly introduce the finite-temperature AFQMC method applied in this work. Then we summarize the entropy evaluation for the half-filling case. In Sec.~\ref{Sec:Ensemble}, we present the computational framework and formulas for thermal entropy in the doped Hubbard model within the grand canonical ensemble, and also derive useful Maxwell relations. Then in Sec.~\ref{Sec:Hellmann}, we present an alternative and independent derivation for the entropy formulas obtained in Sec.~\ref{Sec:Ensemble}. In Sec.~\ref{Sec:Numeric}, we show the test results of entropy in both {2D} and 3D doped Hubbard models from AFQMC simulations. Finally, Sec.~\ref{Sec:Summary} is devoted to the summary of this work, along with discussions on broader applications and further extensions of our computational framework for entropy evaluation. In Appendixes, we discuss the Trotter error of double occupancy and thermal entropy, present derivations for the entropy at certain limits and for several formulas in the main text, and summarize the parameter grids and Monte Carlo sampling details used in the AFQMC calculations.

\section{Model, Method, and entropy evaluation at half filling}
\label{Sec:MMO}

\subsection{The Hubbard model and AFQMC method}
\label{sec:ModelMethod}

We focus on the single-band Fermi-Hubbard model on a {2D} square lattice and a 3D simple cubic lattice to elucidate the computation of thermal entropy. The model Hamiltonian takes the following form:
\begin{equation}\begin{aligned}
\label{eq:Hamiltonian}
\hat{H}
&=\sum_{\mathbf{k}\sigma} \varepsilon_{\mathbf{k}}c_{\mathbf{k}\sigma}^+c_{\mathbf{k}\sigma}^{} + \mu\sum_{\mathbf{i}}(\hat{n}_{\mathbf{i}\uparrow}+\hat{n}_{\mathbf{i}\downarrow}) \\
&\hspace{0.4cm} + U\sum_{\mathbf{i}}\Big(\hat{n}_{\mathbf{i}\uparrow}\hat{n}_{\mathbf{i}\downarrow} - \frac{\hat{n}_{\mathbf{i}\uparrow} + \hat{n}_{\mathbf{i}\downarrow}}{2}\Big),
\end{aligned}\end{equation}
where $\hat{n}_{\mathbf{i}\sigma}=c_{\mathbf{i}\sigma}^+c_{\mathbf{i}\sigma}^{}$ is the density operator, with $\sigma$ ($=\uparrow$ or $\downarrow$) denoting spin and $\mathbf{i}$ as the coordinates of the lattice site. Here we only consider the nearest-neighbor hopping $t$, and take periodic boundary conditions. This leads to the kinetic energy dispersion $\varepsilon_{\mathbf{k}}=-2t(\cos k_x + \cos k_y)$ in {2D}, and $\varepsilon_{\mathbf{k}}=-2t(\cos k_x + \cos k_y + \cos k_z)$ in 3D, where the momentum $k_x,k_y$ (and $k_z$) are defined in units of $2\pi/L$ with $L$ as the linear system size. The total number of lattice sites reads $N_s=L^2$ and $N_s=L^3$ for {2D} and 3D systems, respectively. We define the fermion filling as $n=N_e/N_s$ with $N_e$ as the total number of fermions. Both the repulsive ($U>0$) and attractive ($U<0$) interactions are involved in our calculations. The chemical potential term $\mu$ represents pure doping, since the system with $\mu=0$ is at half filling ($n=1$) due to the particle-hole symmetry. The system is hole doped with $\mu>0$, while the electron doping is associated with $\mu<0$ case. Note that we adopt a positive sign convention for the chemical potential term (i.e., +$\mu$) in the model~(\ref{eq:Hamiltonian}) to maintain consistency with our previous work~\cite{Song2025L,*Song2025B}. Consequently, $\mu\rightarrow+\infty$ corresponds to the empty lattice limit, and the charge compressibility $\kappa$ is defined accordingly as $\kappa=-\partial n(\mu)/\partial\mu$ [see Eq.~(\ref{eq:Chie})].

We employ the {\it numerically exact} finite-temperature AFQMC method~\cite{Blankenbecler1981,Hirsch1983,White1989,Scalettar1991,Assaad2008,YuanYao2019L,YuanYao2019B,Sun2024,Duhao2025,Song2025B,Yuanyao2025} to solve the model~(\ref{eq:Hamiltonian}) and to compute the thermal entropy. While the simulation is free of the fermion sign problem for $U>0$ at half filling and $U<0$ at arbitrary filling~\cite{Congjun2005}, we apply the standard reweighting technique~\cite{White1989,Assaad2008} to treat the sign problem in the repulsive Hubbard model away from half filling. All implementation details of the AFQMC method can be found in our previous work~\cite{Song2025L,*Song2025B}. Here, we briefly outline the key components. We apply the symmetric Trotter-Suzuki decomposition, which generally leads to the Trotter error $O[(\Delta\tau)^2]$ in observables (with $\Delta\tau$ from the discretization of the inverse temperature as $\beta=M\Delta\tau$). We adopt the Hubbard-Stratonovich (HS) transformation into the spin-$\hat{s}^z$ channel with two-component auxiliary fields for the Hubbard interaction~\cite{Hirsch1983,Song2025B}. Our AFQMC simulation also incorporates several algorithmic improvements, including the fast Fourier transform~\cite{Yuanyao2025}, the delayed update~\cite{Sun2024,Duhao2025}, and $\tau$-line global update~\cite{Scalettar1991}. 

\subsection{Computation of thermal entropy at half filling}
\label{Sec:HalfEntropy}

Prior to discussing the doped case, we first summarize the entropy evaluation for the standard Hubbard model at half filling in this subsection. This corresponds to the special case of $\mu=0$ and $n=1$ in the model~(\ref{eq:Hamiltonian}), where the tuning parameters are the interaction strength $U$ and the temperature $T$. The entropy is typically computed either along the $T$ axis at fixed $U$ or along the $U$ axis at fixed $T$. Here we define the entropy density $\eyps=S/N_s$ with $S$ as the total entropy of the system. Both $\eyps$ and $S$ are in units of $k_B$. The calculations involve the energy density $e$ and double occupancy $D$, which are computed as $e=\langle\hat{H}\rangle/N_s$ and $D=N_s^{-1}\sum_{\mathbf{i}}\langle\hat{n}_{\mathbf{i}\uparrow}\hat{n}_{\mathbf{i}\downarrow}\rangle$ in AFQMC simulations.

We begin with the entropy evaluation at fixed $U$ with varying $T$. The conventional approach is to evaluate $\eyps(T)$ via temperature integration~\cite{Dar2007,Ibarra2020,Sushchyev2022,Paiva2010} as
\begin{equation}\begin{aligned}
\label{eq:HalfIntOvT}
\eyps(T) = \ln{4} + \frac{e(T)}{T} - \int_{T}^{\infty} \frac{e(T')}{T^{'2}} \dd T'.
\end{aligned}\end{equation}
Practically, the upper limit of the integral is usually replaced by a high-$T$ cutoff~\cite{Ibarra2020}, which simply neglects the residual tail and may lead to numerical inaccuracy. Moreover, evaluating the integral over a very wide temperature range requires a large number of data points. A possible way to bypass these issues is to reformulate the integral into the inverse temperature axis ($\beta=1/T$) as
\begin{equation}\begin{aligned}
\label{eq:HalfIntOvBeta}
\eyps(\beta) = \ln{4} +\beta e(\beta) - \int_{0}^{\beta} e(\beta') \dd \beta'.
\end{aligned}\end{equation}
This formula has an advantage in handling the high-$T$ regime, since $e(\beta)$ typically varies slowly and smoothly as $T=\infty$. This limit actually corresponds to the atomic limit, where $e(T=\infty)=-U/4$ for the model~(\ref{eq:Hamiltonian}). Therefore, at high temperatures (e.g., $\beta t\le2$), Eq.~(\ref{eq:HalfIntOvBeta}) works well, as only a few data points are usually sufficient to evaluate the integral. However, at intermediate to low temperatures where $\beta$ becomes large, simulations over a wide $\beta$ range are required to accurately evaluate the integral. To overcome this issue, in our previous work~\cite{Song2025B}, we proposed to combine Eqs.~(\ref{eq:HalfIntOvT}) and~(\ref{eq:HalfIntOvBeta}) by dividing the integral into two parts as
\begin{equation}\begin{aligned}
\label{eq:HalfIntBoth}
\eyps(T) = \ln{4} + \frac{e(T)}{T} - \int_{T}^{T_0} \frac{e(T')}{T^{'2}} \dd T' - \int_{0}^{\beta_0} e(\beta') \dd \beta',
\end{aligned}\end{equation}
where $T_0 = 1/\beta_0>T$ denotes an intermediate temperature. In practical calculations, different choices of $T_0$ can be used to cross-check the results for $\eyps(T)$. This hybrid formula separates the calculations into $0\le \beta^{\prime} \le \beta_0$ and $T\le T^{\prime}\le T_0$ regimes, both of which require significantly fewer data points to evaluate the integrals than directly applying Eqs.~(\ref{eq:HalfIntOvT}) or~(\ref{eq:HalfIntOvBeta}). Thus Eq.~(\ref{eq:HalfIntBoth}) can substantially improve the efficiency for the entropy evaluation. Physically, there are two known limits for $\eyps(T)$ in the Hubbard model. The first is $\eyps(T=\infty)=\ln 4$, which corresponds to the atomic-limit result at half filling and is also evident from the expressions in Eqs.~(\ref{eq:HalfIntOvT})--(\ref{eq:HalfIntBoth}). The second is $\eyps(T=0)=0$, which cannot be explicitly derived from the above formulas but holds for the Hubbard model~\cite{EntropyNote}. Connecting these two limits, $\eyps(T)$ simply decays monotonically from $\ln 4$ to zero with lowering temperature.

We now turn to the computation of thermal entropy as a function of $U$ at fixed $T$, denoted as $\eyps(U)$. In principle, $\eyps(U)$ can be obtained by repeating the above procedure to compute $\eyps(T)$ for different values of $U$. However, this approach requires substantial computational effort, as extensive simulations over a wide range of $T$ must be performed for each $U$. To address this challenge, we developed a scheme to directly calculate $\eyps(U)$ in our previous work~\cite{Song2025B}, which only involves simulations at fixed $T$. The idea comes from the Hellmann-Feynman theorem at finite temperatures for the Hubbard model~(\ref{eq:Hamiltonian}) as
\begin{equation}\begin{aligned}
\frac{\partial \gpps}{\partial U} = D - \frac{1}{2},
\end{aligned}\end{equation}
with $\gpps = e - T\eyps(U)$ as the grand potential density. This relation allows for the direct calculation of $\gpps$ via an integral over $U$ as 
\begin{equation}\begin{aligned}
\gpps(U) = \gpps_0 + \int_{0}^{U} D(U') \dd U' - \frac{U}{2},
\end{aligned}\end{equation}
with $\gpps_0=-2(T/N_s)\sum_{\mathbf{k}}{\rm ln}(1+e^{-\beta \varepsilon_{\mathbf{k}}})$ as the grand potential density at $U=0$. Then the following expression for $\eyps(U)$ is evident from the definition of $\gpps$ as
\begin{equation}\begin{aligned}
\label{eq:HalfIntU}
\eyps(U) = \frac{1}{T} \Big[ e(U)-\gpps_0 - \int_{0}^{U} D(U') \dd U' + \frac{U}{2} \Big],
\end{aligned}\end{equation}
which depends on the energy density $e$ and double occupancy $D$. This approach is far more efficient for computing $\eyps(U)$ than repeating the conventional procedure along the $T$ axis at a bunch of $U$ values. It also allows us to carefully track the nonmonotonic $U$ dependence of $\eyps$ at fixed $T$, which provides valuable insights into the metal-insulator crossover in the {2D} and 3D half-filled Hubbard models~\cite{Song2025L,*Song2025B,Lu2026} and the associated Pomeranchuk cooling effect~\cite{Song2025CPL}.

The combined use of Eqs.~(\ref{eq:HalfIntBoth}) and~(\ref{eq:HalfIntU}) enables efficient calculation of thermal entropy in the half-filled Hubbard model. In Eqs.~(\ref{eq:HalfIntOvT})--(\ref{eq:HalfIntBoth}), the energy density $e$ can be replaced by a simplified analog $\widetilde{e}=N_s^{-1}\langle\hat{H}_K\rangle+UD$, where $\hat{H}_K=\sum_{\mathbf{k}\sigma}\varepsilon_{\mathbf{k}}c_{\mathbf{k}\sigma}^+c_{\mathbf{k}\sigma}^{}$ is the kinetic Hamiltonian. It is evident that the extra energy $\Delta e=e-\widetilde{e}=-U/2$ at half filling cancels out in all the formulas. Accordingly, the expression in Eq.~(\ref{eq:HalfIntU}) can be reformulated as
\begin{equation}\begin{aligned}
\label{eq:HalfIntUDou}
\eyps(U) = \frac{1}{T} \Big[ \widetilde{e}(U)-\gpps_0 - \int_{0}^{U} D(U') \dd U'\Big].
\end{aligned}\end{equation}
Regarding the computation of the integrals in Eqs.~(\ref{eq:HalfIntBoth}) and~(\ref{eq:HalfIntU}), we suggest first performing fitting (e.g., cubic spline) for numerical data of the integrands [as $e(T^{\prime})$ and $e(\beta^{\prime})$ in Eq.~(\ref{eq:HalfIntBoth}), and $D(U^{\prime})$ in Eq.~(\ref{eq:HalfIntU})], and then evaluate the integral analytically using the fitting curve. The uncertainty of the entropy can be estimated via the bootstrapping technique (see Appendix C in Ref.~\cite{Song2024}). The same procedure is also used in numerically evaluating the integrals involved in the next section for the doped case.

\section{Entropy evaluation away from half filling and Maxwell relations}
\label{Sec:Ensemble}

In this section, we concentrate on our unified framework for computing the thermal entropy for the doped Hubbard model. Compared with the half-filling case, the doping introduces the additional parameter $\mu$, which is the characteristic variable of the grand canonical ensemble. The framework is formulated based on the fundamental quantity of grand potential.

We first clarify the terminology used below. In previous literature, the thermodynamic potential in the grand canonical ensemble is sometimes loosely called the {\it free energy}. More precisely, however, {\it free energy} usually refers to the Helmholtz free energy in the canonical ensemble, while the corresponding quantity in the grand canonical ensemble is the {\it grand potential}. In this work, since all calculations are carried out in the grand canonical ensemble, we use the term {\it grand potential} throughout. Correspondingly, the grand potential density is regarded as a function of temperature $T$, interaction strength $U$, and chemical potential $\mu$. With our convention for the Hubbard Hamiltonian, the chemical-potential term is included explicitly in $\hat{H}$, and thus the grand potential density is written as $\gpps(T,U,\mu)=e-T\eyps$, where $\eyps=S/N_s$ is the entropy density and $e=\langle\hat{H}\rangle/N_s=e_k+Uh_I+\mu n$ is the total energy density including the chemical-potential contribution. Here $e_k=N_s^{-1}\langle\hat{H}_K\rangle$ is the kinetic energy density, while $h_I=\langle\hat{H}_I\rangle/N_s$ with $\hat{H}_I=\sum_{\mathbf{i}}[\hat{n}_{\mathbf{i}\uparrow}\hat{n}_{\mathbf{i}\downarrow}-(\hat{n}_{\mathbf{i}\uparrow}+\hat{n}_{\mathbf{i}\downarrow})/2]$. Therefore $\gpps(T,U,\mu)$ can be expressed explicitly as
\begin{equation}\begin{aligned}
\label{eq:GPdensity}
\gpps = e_{k} +Uh_{I}- T\eyps + \mu n,
\end{aligned}\end{equation}
where $\eyps=S/N_s$ is the entropy density.

Based on the fact that the natural variables in $\gpps$ are $T$, $U$, and $\mu$, the total differential of $\gpps$ should be physically written as
\begin{equation}\begin{aligned}
\label{eq:TotDiff1}
\dd{\gpps}=-\eyps \dd{T}+h_{I} \dd{U}+n\dd{\mu}.
\end{aligned}\end{equation}
Moreover, from Eq.~(\ref{eq:GPdensity}), we can calculate the total differential of $\gpps$ in a totally mathematical manner as
\begin{equation}\begin{aligned}
\label{eq:TotDiff2}
\dd{\gpps}= \dd{e_k} &+ h_I\dd{U} + U\dd{h_I} - T\dd{\eyps} - \eyps\dd{T} \\
&{\hspace{0.9cm}} + \mu\dd{n} + n\dd{\mu}.
\end{aligned}\end{equation}
By comparing Eqs.~(\ref{eq:TotDiff1}) and~(\ref{eq:TotDiff2}), we reach the total differential of the entropy density as
\begin{equation}\begin{aligned}
\label{eq:TotDiffEntropy}
\dd \eyps = \frac{\dd{e_k} + U\dd{h_{I}} + \mu \dd{n}}{T}.
\end{aligned}\end{equation}
This relation clearly provides the practical routes to compute $\eyps$ along different pathways by integrating specific parameter trajectories. 

To systematically evaluate the entropy along physically relevant pathways, we consider four representative paths in the $(U,T,\mu)$ parameter space of the model~(\ref{eq:Hamiltonian}), each involving the variation of a single control parameter while keeping the others fixed. Specifically, we concentrate on the following four paths: (a) varying $T$ with fixed $U$ and $n$ [see Eq.~(\ref{eq:fixednT})]; (b) varying $T$ with fixed $U$ and $\mu$ [see Eq.~(\ref{eq:fixedmT})]; (c) varying $\mu$ with fixed $U$ and $T$ [see Eq.~(\ref{eq:changem})]; (d) varying $U$ with fixed $T$ and $n$ [see Eqs.~(\ref{eq:changeUfixTn}) and~(\ref{eq:changeUfixTn1})]. A concrete illustration of these four paths, based on the simulation parameters (see Sec.~\ref{Sec:Testin3D}), is provided in Fig.~\ref{fig:Fig01Paths}. These basically cover the whole parameter space of the doped Hubbard model. For clarity and later convenience for use, we define two different kinds of energy density
\begin{equation}\begin{aligned}
\label{eq:DefEng}
e_{\rm in} &= e_k + U h_{I}, \\
e_{\rm tot} &= e_k+U h_{I} + \mu n,
\end{aligned}\end{equation}
where $e_{\rm tot}$ includes the chemical potential term. Both $e_{\rm in}$ and $e_{\rm tot}$ will be used in the following calculations for the entropy.

In the following, we present the detailed derivations for the entropy density $\eyps$ based on Eq.~(\ref{eq:TotDiffEntropy}) along the above four paths in Secs.~\ref{Sec:FixUn}--\ref{Sec:FixTn}, and discuss the Maxwell relations connecting the entropy to other thermal quantities in Sec.~\ref{Sec:Maxwell}. We also elucidate an efficient technique to tune the chemical potential $\mu$ to reach a fixed fermion filling in the context of the grand canonical ensemble in Sec.~\ref{Sec:TuneMu}.

\subsection{Varying \texorpdfstring{$T$}{} with fixed \texorpdfstring{$U$}{} and \texorpdfstring{$n$}{}}
\label{Sec:FixUn}

This is the standard path that is usually applied in quantum many-body simulations for the doped Hubbard model~\cite{YuanYao2019B,Qiaoyi2026}. In the grand canonical ensemble calculations, the fixed fermion filling $n$ is typically reached via tuning $\mu$ by hand (see Sec.~\ref{Sec:TuneMu}). 

Starting from Eq.~(\ref{eq:TotDiffEntropy}), the fixed $n$ means $\dd n=0$, and fixed $U$ leads to $\dd{e_k} + U\dd{h_{I}}=\dd{e_{\rm in}}$. This leads to the simplified total differential $\dd{\eyps}=\dd{e_{\rm in}}/T$, which can be easily transformed to an integral form as
\begin{equation}\begin{aligned}
\label{eq:fixedn_rough}
\eyps_n(\infty) - \eyps(T) = \int_{T}^{\infty} \frac{\dd{e_{\rm in}(T^{\prime})}}{T^{\prime}},
\end{aligned}\end{equation}
where $\eyps_n(\infty) = \ln 4 - n\ln n - (2-n)\ln(2-n)$ is the entropy density at $T=\infty$ with the fixed filling $n$ (see Appendix~\ref{Sec:A2Entropy}). Note $\eyps_n(\infty)$ reduces to the half-filling value of $\ln 4$ at $n=1$. To make Eq.~(\ref{eq:fixedn_rough}) numerically tractable, we perform the integration by parts and arrive at
\begin{equation}\begin{aligned}
\label{eq:fixedn}
\eyps(T)= \eyps_n(\infty) &+ \frac{e_{\rm in}(T)}{T} - \int_{T}^{\infty}\frac{e_{\rm in}(T^{\prime})}{{T^{\prime}}^2} \dd T^{\prime}.
\end{aligned}\end{equation}
We then apply the same operation as that in Sec.~\ref{Sec:HalfEntropy} to deal with the upper limit $T=\infty$ in the integral, by dividing it into two parts separated by an intermediate temperature $T_0>T$ (and $\beta_0 = 1/T_0$) as
\begin{equation}\begin{aligned}
\label{eq:fixednT}
\eyps(T) = \eyps_n(\infty) &+ \frac{e_{\rm in}(T)}{T} - \int_T^{T_0}\frac{e_{\rm in}(T^{\prime})}{{T^{\prime}}^2} \dd T^{\prime} \\
&- \int_0^{\beta_0} e_{\rm in}(\beta^{\prime}) \dd \beta^{\prime}.
\end{aligned}\end{equation}
At the special filling $n=1$, this expression clearly degenerates into Eq.~(\ref{eq:HalfIntBoth}) for the half-filling case. In Eqs.~(\ref{eq:fixedn}) and~(\ref{eq:fixednT}), the energy $e_{\rm in}$ can be replaced by $\widetilde{e}_{\rm in}=e_k+UD$, as the extra term $e_{\rm in}-\widetilde{e}_{\rm in}=-Un/2$ cancels out.

An alternative choice for formulating the integration in Eq.~(\ref{eq:fixedn_rough}) is to integrate from $T^{\prime}=0$ to $T^{\prime}=T$, which can benefit from $\eyps_n(T=0)=0$. However, it requires the results of $e_{\rm in}(T)$ in the low-$T$ regime, which is typically more challenging for quantum many-body numerical methods, especially for the AFQMC method due to the fermion sign problem.

\subsection{Varying \texorpdfstring{$T$}{} with fixed \texorpdfstring{$U$}{} and \texorpdfstring{$\mu$}{}}
\label{Sec:FixUmu}

In addition to a fixed-$n$ trajectory, another frequently studied path involves varying temperature $T$ with fixed $U$ and $\mu$. This path can be applied in quantum many-body calculations when fixing $n$ requires substantially greater computational effort. 

With fixed $\mu$, the fermion filling $n$ changes as $T$ decreases, implying $\dd{n}\ne0$. Consequently, the chemical potential term $\mu\dd{n}$ in Eq.~(\ref{eq:TotDiffEntropy}) contributes to total differential $\dd{\eyps}$, which is the major difference from the fixed-$n$ path as discussed in Sec.~\ref{Sec:FixUn}. Together with fixed $U$, it leads to $\dd{e_k} + U\dd{h_{I}} + \mu \dd{n}=\dd{e_{\rm tot}}$, resulting in the total differential $\dd{\eyps}=\dd{e_{\rm tot}}/T$. Hence, the entropy can be computed as
\begin{equation}\begin{aligned}
\label{eq:fixedmT_rough}
\eyps_{\mu}(\infty) - \eyps(T) = \int_{T}^{\infty} \frac{\dd{e_{\rm tot}(T^{\prime})}}{T^{\prime}},
\end{aligned}\end{equation}
where $\eyps_{\mu}(\infty) = \ln 4$ is the entropy density at $T=\infty$ with the fixed $\mu$. As $T\to\infty$, the system approaches the atomic limit, which holds $n=1$ for any finite $\mu$ (see Appendix~\ref{Sec:A2Entropy}) and thus yields $\eyps_{\mu}(\infty) = \ln 4$. Then by applying integration by parts to Eq.~(\ref{eq:fixedmT_rough}) and introducing an intermediate temperature $T_0>T$ (and $\beta_0 = 1/T_0$), we reach the final expression for $\eyps(T)$ as
\begin{equation}\begin{aligned}
\label{eq:fixedmT}
\eyps(T) = \ln 4 &+ \frac{e_{\rm tot}(T)}{T} - \int_T^{T_0}\frac{e_{\rm tot}(T^{\prime})}{{T^{\prime}}^2} \dd T^{\prime} \\
&- \int_0^{\beta_0} e_{\rm tot}(\beta^{\prime}) \dd \beta^{\prime},
\end{aligned}\end{equation}
which shares the same form as Eq.~(\ref{eq:fixednT}) for the fixed-$n$ path, the only difference being the replacement of $e_{\rm in}(T)$ by $e_{\rm tot}(T)$. At half filling with $\mu=0$ and $n=1$, the relation $e_{\rm tot}=e_{\rm in}$ holds, and hence Eqs.~(\ref{eq:fixedmT}) and~(\ref{eq:fixednT}) become identical, which both degenerate into Eq.~(\ref{eq:HalfIntBoth}) as expected.

With $\mu\ne0$, we can obtain two types of specific heat along the trajectories of fixed $n$ and fixed $\mu$ applied to calculate the entropy. The specific heat can be computed as
\begin{equation}\begin{aligned}
C_X = T\Big(\frac{\partial \eyps}{\partial T}\Big)_X,
\end{aligned}\end{equation}
with $X=n$ or $X=\mu$ denoting the fixed quantity of the path. Using the expressions for $\eyps(T)$ in Eqs.~(\ref{eq:fixednT}) and~(\ref{eq:fixedmT}) along the two paths, we obtain
\begin{equation}\begin{aligned}
\label{eq:SpecHeat}
&C_{n} = T\Big(\frac{\partial \eyps}{\partial T}\Big)_n = \Big(\frac{\partial e_{\mathrm{in}}}{\partial T}\Big)_n, \\
&C_\mu = T\Big(\frac{\partial \eyps}{\partial T}\Big)_\mu = \Big(\frac{\partial e_{\mathrm{tot}}}{\partial T}\Big)_\mu.
\end{aligned}\end{equation}
These two formulas can be used to compute the specific heat along the corresponding paths to characterize the physical properties of the system.

\subsection{Varying \texorpdfstring{$\mu$}{} with fixed \texorpdfstring{$U$}{} and \texorpdfstring{$T$}{}}
\label{Sec:FixUT}

As a complementary perspective on varying $T$ schemes in Secs.~\ref{Sec:FixUn} and~\ref{Sec:FixUmu}, we can also evaluate the entropy as a function of $\mu$ (namely, doping or $n$) or $U$ at fixed $T$. Such a formulation is particularly useful in characterizing the doping physics in the model. We first focus on the path of varying $\mu$ with fixed $U$ and $T$.

In Eq.~(\ref{eq:TotDiffEntropy}), fixing $U$ means $\dd{e_k} + U\dd{h_{I}}=\dd{e_{\rm in}}$, and results in the total differential $\dd{\eyps}=(\dd{e_{\rm in}}+\mu\dd{n})/T$. We can similarly perform integration on both sides from a specific reference point $\mu^{\prime}=\mu_0$ to $\mu^{\prime}=\mu$, leading to
\begin{equation}\begin{aligned}
\eyps(\mu) - \eyps(\mu_0) = \frac{1}{T}\Big[ e_{\mathrm{in}}(\mu) - e_{\mathrm{in}}(\mu_0) +\int_{\mu_0}^{\mu} \mu^{\prime} \dd{n(\mu^{\prime})} \Big].
\end{aligned}\end{equation}
Applying the integration by parts for the integral, we can reach its simplified form as
\begin{equation}\begin{aligned}
\label{eq:changem_Raw}
\eyps(\mu) = \eyps(\mu_0) + \frac{1}{T}\Big[ e_{\mathrm{tot}}(\mu) - e_{\mathrm{tot}}(\mu_0) -\int_{\mu_0}^{\mu} n(\mu^{\prime}) \dd{\mu^{\prime}} \Big],
\end{aligned}\end{equation}
where we apply the relation $e_{\rm tot}(\mu)=e_{\rm in}(\mu)+\mu n(\mu)$ from Eq.~(\ref{eq:DefEng}). In practical calculations, the reference $\mu_0$ can be chosen at a point where the numerical simulations are relatively straightforward. For the Hubbard model~(\ref{eq:Hamiltonian}), the half-filling point with $\mu_0=0$ is obviously the best choice, and the corresponding $\eyps(\mu_0=0)$ can be computed via either Eq.~(\ref{eq:HalfIntBoth}) or Eq.~(\ref{eq:HalfIntU}). Accordingly, Eq.~(\ref{eq:changem_Raw}) with $\mu_0=0$ can be written as
\begin{equation}\begin{aligned}
\label{eq:changem}
\eyps(\mu) = \eyps(0) + \frac{1}{T}\Big[ e_{\mathrm{tot}}(\mu) - e_{\mathrm{tot}}(0) -\int_{0}^{\mu} n(\mu^{\prime}) \dd{\mu^{\prime}} \Big],
\end{aligned}\end{equation}
where $\eyps(0)$ and $e_{\mathrm{tot}}(0)$ are the entropy density and total energy density, respectively, at $\mu=0$ (half filling). Note that the energy $e_{\mathrm{tot}}$ instead of $e_{\mathrm{in}}$ appears in Eq.~(\ref{eq:changem}), since the fermion filling $n$ changes along the trajectory of varying $\mu$. Another special choice for $\mu_0$ is $\mu_0=+\infty$, at which the fermion density $n=0$ in the model~(\ref{eq:Hamiltonian}). However, this requires wide $\mu$-sweep simulations to evaluate the integral, which consumes more computational effort.

Although $\eyps(0)$ in Eq.~(\ref{eq:changem}) must be computed independently, typically through the integration along the $T$ or $U$ axis as described in Sec.~\ref{Sec:MMO}, it contributes only an additive constant when analyzing the entropy differences across fillings. In practical applications, this constant is irrelevant when only the relative variation or the shape of the entropy curve as a function of doping is of interest. This path therefore provides a flexible and efficient route to access entropy trends in the doped Hubbard model, particularly in regimes where fixing the fermion filling directly is computationally demanding. For example, at intermediate to low temperatures where the fermion sign problem becomes significant in AFQMC simulations, this scheme is particularly useful since the fermion density $n(\mu)$ can typically be obtained with high precision.

\subsection{Varying \texorpdfstring{$U$}{} with fixed \texorpdfstring{$T$}{} and \texorpdfstring{$n$}{}}
\label{Sec:FixTn}

The last computational scheme for the entropy density $\eyps$ is along the $U$ axis with fixed $T$ and $n$. It enables us to study the $U$ dependence of $\eyps$, which plays a significant role in characterizing the metal-to-insulator crossover in the Hubbard models~\cite{Song2025L,*Song2025B,Lu2026}.

Based on Eq.~(\ref{eq:TotDiffEntropy}), the fixed $n$ means $\dd n=0$, leading to the total differential $\dd{\eyps}=(\dd{e_k}+U\dd{h_I})/T$. Note that $h_I$ also changes with varying $U$. Then we compute $\eyps$ by performing the integration from $U^{\prime}=0$ to $U^{\prime}=U$, yielding
\begin{equation}\begin{aligned}
\label{eq:changeUIni0}
\eyps(U) - \eyps(0) = \frac{1}{T}\Big[ e_k(U) - e_k(0) + \int_{0}^{U} U^{\prime} \dd{h_{I}(U^{\prime})} \Big].
\end{aligned}\end{equation}
Here, $\eyps(0)=\eyps_0$ is the noninteracting ($U=0$) entropy density with fixed $T$ and $n$, which takes the explicit form (see Appendix~\ref{Sec:A2Entropy})
\begin{equation}\begin{aligned}
\label{eq:U0Exact}
\eyps_0 = -2N_s^{-1}\sum_{\mathbf{k}}\big[ f_{\mathbf{k}} \ln f_{\mathbf{k}} + (1-f_{\mathbf{k}})\ln(1-f_{\mathbf{k}})\big],
\end{aligned}\end{equation}
with $f_{\mathbf{k}} =1/[e^{\beta(\varepsilon_{\mathbf{k}}+\mu)}+1]$ as the Fermi-Dirac distribution function. Via the integration by parts for the integral in Eq.~(\ref{eq:changeUIni0}), the expression can be simplified as
\begin{equation}\begin{aligned}
\label{eq:changeUIni1}
\eyps(U) - \eyps_0 = \frac{1}{T} \Big[ e_{\mathrm{in}}(U) - e_{\mathrm{in}}(0) - \int_{0}^{U} h_{I}(U^{\prime}) \dd{U^{\prime}} \Big],
\end{aligned}\end{equation}
which incorporates the relation $e_{\rm in}(U) = e_k(U) + U h_{I}(U)$ from Eq.~(\ref{eq:DefEng}). Moreover, using the equality  $h_{I}(U^{\prime})=D(U^{\prime})-n/2$, Eq.~(\ref{eq:changeUIni1}) can be rewritten as
\begin{equation}\begin{aligned}
\label{eq:changeUfixTn}
&\eyps(U) = \eyps_0 \\
&+ \frac{1}{T} \Big[ e_{\mathrm{in}}(U) - e_{\mathrm{in}}(0) - \int_{0}^{U} D(U^{\prime}) \dd{U^{\prime}} + \frac{Un}{2} \Big].
\end{aligned}\end{equation}
Considering the fixed $n$, we can reformulate this equality using $\widetilde{e}_{\rm in}(U)=e_k(U)+UD(U)=e_{\rm in}(U)+Un/2$ as
\begin{equation}\begin{aligned}
\label{eq:changeUfixTn1}
\eyps(U) = \eyps_0 + \frac{1}{T} \Big[ \widetilde{e}_{\rm in}(U) - \widetilde{e}_{\rm in}(0) - \int_{0}^{U} D(U^{\prime}) \dd{U^{\prime}} \Big].
\end{aligned}\end{equation}
It is evident that, at $n=1$, Eqs.~(\ref{eq:changeUfixTn}) and~(\ref{eq:changeUfixTn1}) become identical to Eqs.~(\ref{eq:HalfIntU}) and~(\ref{eq:HalfIntUDou}) for the half-filling case, respectively, based on the relations $\gpps_0=e_k - T\eyps_0$ and $e_{\rm in}(0)=\widetilde{e}_{\rm in}(0)=e_k(U=0)$.

Typically, double occupancy $D(U)$ is a very smooth function of $U$ with fixed $T$ and $n$. As a result, we can accurately evaluate the integral in Eq.~(\ref{eq:changeUfixTn1}) and accordingly obtain the full $\eyps(U)$ curve, using a limited number of data points from numerical simulations.

\subsection{Practical comparison of the entropy-evaluation schemes}
\label{Sec:PathComparison}

\begin{table*}[t]
\caption{
Practical comparison of the four entropy-evaluation schemes developed in this work. For each scheme, we outline the integration path, primary observable, computational cost, statistical robustness, and recommended usage. The energy densities $e_{\rm in}$ and $e_{\rm tot}$ are defined in Eq.~(\ref{eq:DefEng}), $n$ is the fermion filling, and $D$ is double occupancy. The doping is defined as $\delta=1-n$.
}
\label{tab:path_comparison}
\centering
\footnotesize
\renewcommand{\arraystretch}{1.20}
\newcommand{\ccell}[1]{%
\begin{tabular}[c]{@{}c@{}}#1\end{tabular}%
}
\begin{tabular*}{\textwidth}{@{\extracolsep{\fill}}c c c c c c}
\hline\hline
Scheme &
Path &
Primary observable &
Computational cost &
Statistical robustness &
Recommended usage \\
\hline
		
\ccell{(a), Eq.~(\ref{eq:fixednT})} &
\ccell{Vary $T$ \\ at fixed $U,n$} &
\ccell{$e_{\rm in}(T)$} &
\ccell{Moderate} &
\ccell{Moderate} &
\ccell{Scanning $T$} \\
\hline
		
\ccell{(b), Eq.~(\ref{eq:fixedmT})} &
\ccell{Vary $T$ \\ at fixed $U,\mu$} &
\ccell{$e_{\rm tot}(T)$} &
\ccell{Moderate} &
\ccell{Moderate} &
\ccell{Scanning $T$} \\
\hline
		
\ccell{(c), Eq.~(\ref{eq:changem})} &
\ccell{Vary $\mu$ \\ at fixed $U,T$} &
\ccell{$n(\mu)$ and $e_{\rm tot}(\mu)$} &
\ccell{Low} &
\ccell{High} &
\ccell{Scanning doping $\delta$} \\
\hline
		
\ccell{(d), Eq.~(\ref{eq:changeUfixTn})} &
\ccell{Vary $U$ \\ at fixed $T,n$} &
\ccell{$D(U)$ and $e_{\rm in}(U)$} &
\ccell{Low} &
\ccell{High} &
\ccell{Scanning $U$} \\
		
\hline\hline
\end{tabular*}
\end{table*}

The four entropy-evaluation schemes introduced above (in Secs.~\ref{Sec:FixUn}--\ref{Sec:FixTn}) provide complementary approaches for computing the thermal entropy along different thermodynamic integration paths. Practically, however, their performance depends on the target parameter regime. In this subsection, we compare the efficiency of these schemes in practical simulations. In Table~\ref{tab:path_comparison}, we summarize the integration path, primary observable, computational cost, statistical robustness, and recommended applications for each scheme.

Schemes (a) and (b) are the natural choices for studying the temperature dependence of the entropy at fixed filling ($n$) and fixed chemical potential ($\mu$), respectively. They correspond to Eqs.~(\ref{eq:fixednT}) and~(\ref{eq:fixedmT}), and use the hybrid $T$-$\beta$ integration to avoid the high temperature cutoff. The corresponding integrands are the energy densities $e_{\rm in}$ and $e_{\rm tot}$ [see Eq.~(\ref{eq:DefEng})] for the schemes (a) and (b), respectively. Both quantities are generally smooth functions of temperature and can be evaluated accurately. Thus, in practical calculations using Eqs.~(\ref{eq:fixednT}) and~(\ref{eq:fixedmT}), only a few data points are typically required for the high-temperature $\beta$ integral, together with a reasonable number of data points for the low-temperature $T$ integral, thereby resulting in moderate computational cost. However, the relatively long integration paths (especially at low $T$) may result in modest error accumulation, and approaching $T=0$ the small entropy can make the cancellation in the formulas more delicate. Overall, the two schemes provide stable integrands for the thermodynamic integration and exhibit moderate statistical robustness. For example, in our previous studies of the half-filled 3D Hubbard model~\cite{Song2025L,*Song2025B,Song2025CPL}, these schemes produced stable entropy results down to $T/t=0.1$. In practice, scheme (a) is more widely adopted because the fermion filling $n$ is typically the physically relevant control parameter, whereas scheme (b) is mainly useful for simulations along fixed-$\mu$ trajectories or for qualitative comparisons.

Schemes (c) and (d) can conveniently evaluate the doping ($\delta=1-n$) and interaction ($U$) dependences of thermal entropy, as revealed by Eqs.~(\ref{eq:changem}) and~(\ref{eq:changeUfixTn}). In scheme (c), the integrand is fermion filling $n(\mu)$, together with the endpoint values of $e_{\rm tot}$, while in scheme (d), the corresponding quantities are the double occupancy $D(U)$ and the endpoint energy density $e_{\rm in}$. The numerical data for $n(\mu)$ and $D(U)$ are typically smooth and can be obtained with high precision~\cite{Lu2026B}. These two schemes involve shorter integration paths over $\mu$ or $U$, compared with the hybrid integration over a wide parameter range in schemes (a) and (b). Moreover, since only numerical simulations at fixed $T/t$ are required in schemes (c) and (d), schemes (c) and (d) feature relatively low computational cost. These features reduce error accumulation and make the corresponding integration more stable against statistical fluctuations. These two schemes are also efficient when computing thermal entropy at isolated parameter points.

The above comparison in Table~\ref{tab:path_comparison} and associated discussions assume that the underlying observables can be obtained with reliable precision using various quantum many-body approaches. Nevertheless, for AFQMC simulations, the situation can be affected by the fermion sign problem. When the sign problem is not too severe (i.e., the sign average $\langle\mathcal{S}\rangle\ge0.1$), all four schemes remain effective since the energies $e_{\rm in}$ and $e_{\rm tot}$, fermion filling $n$, and double occupancy $D$ can all be computed with sufficient precision. Within this regime of a controllable sign problem, schemes (c) and (d) are generally more favorable because $n(\mu)$ and $D(U)$ typically have smaller statistical uncertainties than $e_{\rm in}$ and $e_{\rm tot}$. Hence, using $n(\mu)$ and $D(U)$ as integrands leads to less error accumulation during the thermodynamic integration than using $e_{\rm in}$ and $e_{\rm tot}$ in schemes (a) and (b). For instance, when computing the entropy as a function of the hole doping $\delta=1-n$ at fixed $T/t$ in the doped Hubbard model, scheme (c) is much more efficient than the varying-$T$ schemes (a) and (b) (see Sec.~\ref{Sec:PhysicsIn{2D}}). On the other hand, when the sign problem becomes severe, the statistical uncertainties of all relevant observables increase substantially, making it difficult for any of the four schemes to produce high-precision entropy estimates.

\subsection{Maxwell relations}
\label{Sec:Maxwell}

In the previous sections, we have derived formulas for the entropy along four different paths in the parameter space of the doped Hubbard model, which together cover all relevant simulation scenarios. With the entropy formulas, here we focus on the Maxwell relations that connect thermal entropy with other physical quantities in the model. In the grand canonical ensemble, Maxwell relations follow from the equality of mixed second-order derivatives of the grand potential. In the procedure, we need to carefully distinguish the fixed-$\mu$ and fixed-$n$ conditions. 

From the grand potential density $\gpps(U,T,\mu)$ in Eq.~(\ref{eq:GPdensity}), we can readily obtain its first-order derivatives as
\begin{equation}\begin{aligned}
\frac{\partial \gpps}{\partial T} = -\eyps,\quad
\frac{\partial \gpps}{\partial U} = h_I,\quad
\frac{\partial \gpps}{\partial \mu} = n,
\end{aligned}\end{equation}
which then yields the mixed second-order derivatives
\begin{subequations}\begin{align}
\Big(\frac{\partial \eyps}{\partial U}\Big)_{\mu,T} &= -\Big(\frac{\partial h_{I}}{\partial T}\Big)_{\mu,U}, \label{eq:Maxwella} \\
\Big(\frac{\partial \eyps}{\partial \mu}\Big)_{U,T} &= -\Big(\frac{\partial n}{\partial T}\Big)_{U,\mu}, \label{eq:Maxwellb}\\
\Big(\frac{\partial h_{I}}{\partial \mu}\Big)_{T,U} &= +\Big(\frac{\partial n}{\partial U}\Big)_{T,\mu}, \label{eq:Maxwellc}
\end{align}\end{subequations}
where the subscripts (such as ``$\mu,T$'' and ``$\mu,U$'') denote the condition of fixed natural variables. Note that, in our model Hamiltonian~(\ref{eq:Hamiltonian}), we write the chemical potential term $+\mu\sum_{\mathbf{i}}(\hat{n}_{\mathbf{i\uparrow}}+\hat{n}_{\mathbf{i\downarrow}})$ as a pure hole-doping term with $\mu>0$. Equations~(\ref{eq:Maxwella})--(\ref{eq:Maxwellc}) are the Maxwell relations for the fixed-$\mu$ calculations. They already have practical applications. {\it First}, fixing $\mu=0$ in the model~(\ref{eq:Hamiltonian}) leads to $n=1$ for arbitrary $U$ and $T$, as the half-filling case. This yields $h_I=D-1/2$, and hence Eq.~(\ref{eq:Maxwella}) can be rewritten as $(\partial\eyps/\partial U)_{T}=-(\partial D/\partial T)_{U}$, the well-known Maxwell relation in the half-filled Hubbard model~\cite{Song2025B,Lu2026}. {\it Second}, Eq.~(\ref{eq:Maxwellb}) means that, at fixed $U$, the variation of $\eyps$ with $\mu$ [as $(\partial\eyps/\partial\mu)_{U,T}$] is opposite to the change of fermion density $n$ versus $T$ [as $(\partial n/\partial T)_{U,\mu}$]. Consequently, the extrema of $\eyps(\mu)$ and $n(T)$ occur at the same $(T,\mu)$ point. As a special case, $\mu=0$ in the model~(\ref{eq:Hamiltonian}) corresponds to the half filling, and hence leads to $(\partial n/\partial T)_{U,\mu=0}=0$ and subsequently $(\partial\eyps/\partial\mu)_{U,T}=0$, meaning that $\mu=0$ is an extremum of the $\eyps(\mu)$ function at fixed $T$ and $U$. This can also be revealed by directly computing $(\partial\eyps/\partial\mu)_{U,T}$ via Eq.~(\ref{eq:changem}), which yields
\begin{equation}\begin{aligned}
\label{eq:ParSParMu}
\Big(\frac{\partial\eyps}{\partial\mu}\Big)_{U,T}
&= \frac{1}{T}\Big[ \frac{\partial e_{\mathrm{tot}}(\mu)}{\partial\mu} - n(\mu) \Big]_{U,T} \\
&= \frac{1}{T}\Big[ \frac{\partial e_{\mathrm{in}}(\mu)}{\partial\mu} + \mu\frac{\partial n(\mu)}{\partial\mu}\Big]_{U,T},
\end{aligned}\end{equation}
based on $e_{\mathrm{tot}}(\mu)=e_{\mathrm{in}}(\mu)+\mu n(\mu)$ from Eq.~(\ref{eq:DefEng}). Moreover, the hole-doped (with $\mu>0$) and electron-doped (with $\mu<0$) situations of the model~(\ref{eq:Hamiltonian}) are connected via particle-hole transformations $c_{\mathbf{i}\sigma}^+\to(-1)^{\mathbf{i}}c_{\mathbf{i}\sigma}^{},c_{\mathbf{i}\sigma}^{}\to(-1)^{\mathbf{i}}c_{\mathbf{i}\sigma}^+$, which leads to $n(-\mu)=2-n(\mu)$ and $e_{\mathrm{in}}(-\mu)=e_{\mathrm{in}}(\mu)$. Both $e_{\mathrm{in}}(\mu)$ and $n(\mu)$ are expected to be analytic functions near $\mu=0$ (in the absence of phase transitions), resulting in $(\partial e_{\mathrm{in}}/\partial \mu)_{\mu=0}=0$, and consequently $(\partial \eyps/\partial \mu)_{U,T}=0$ at $\mu=0$. Equation~(\ref{eq:ParSParMu}) also indicates that $(\partial \eyps/\partial \mu)_{U,T}$ has opposite signs but equal magnitudes at $+\mu$ and $-\mu$ points, implying $\eyps(\mu)=\eyps(-\mu)$ in the model~(\ref{eq:Hamiltonian}) for fixed $U$ and $T$. A direct proof of $\eyps(\mu)=\eyps(-\mu)$ can also be achieved via Eq.~(\ref{eq:changem}) (see Appendix~\ref{Sec:A2Entropy}). Furthermore, Eq.~(\ref{eq:ParSParMu}) can be used to numerically compute the extremum of $\eyps(\mu)$ via $(\partial\eyps/\partial\mu)_{U,T}=0$. {\it Third}, taking an additional $\mu$ derivative for Eq.~(\ref{eq:Maxwellb}) can lead to a new equality
\begin{equation}\begin{aligned}
\label{eq:ParSParMu2nd}
\Big(\frac{\partial^2 \eyps}{\partial \mu^2}\Big)_{U,T}
= \Big(\frac{\partial\kappa}{\partial T}\Big)_{U,\mu},
\end{aligned}\end{equation}
in which $\kappa=-\partial n/\partial\mu$ is the charge compressibility, and the equality of mixed partial derivatives is applied. Typically, $\partial\kappa/\partial T <0$ and $\partial\kappa/\partial T >0$ can qualitatively separate the metallic and insulating behaviors~\cite{Kim2020}. Based on this physical interpretation and Eq.~(\ref{eq:ParSParMu2nd}), the condition $\partial^2 \eyps/\partial \mu^2=0$ is identified as the crossover from metallic behavior to a non-Fermi-liquid state in Ref.~\cite{Lenihan2021}. 

While the fixed-$\mu$ case discussed above follows directly from the natural variables of the grand canonical ensemble, many theoretical and experimental situations instead involve fixing the fermion filling $n$. Therefore we now turn to the Maxwell relations under the fixed-$n$ condition for the doped Hubbard model. These relations can be obtained by enforcing the total differential conditions along paths of constant filling. {\it First}, at fixed $U$ and $n$, the total differential $\dd n$ is given by
\begin{equation}\begin{aligned}
\label{eq:TotDiffn}
\dd n = \Big(\frac{\partial n}{\partial\mu}\Big)_{U,T}\dd\mu + \Big(\frac{\partial n}{\partial T}\Big)_{U,\mu}\dd T = 0,
\end{aligned}\end{equation}
and it subsequently yields a Maxwell relation
\begin{equation}\begin{aligned}
\label{eq:Maxwell000}
\Big(\frac{\partial \mu}{\partial T}\Big)_{U,n}
= -\frac{\big(\frac{\partial n}{\partial T}\big)_{U,\mu}}
{\big(\frac{\partial n}{\partial \mu}\big)_{U,T}}
= \Big(\frac{\partial \eyps}{\partial n}\Big)_{U,T},
\end{aligned}\end{equation}
in which Eq.~(\ref{eq:Maxwellb}) is used in the second equality. This relation was also discussed in Ref.~\cite{Khatami2009,Khatami2011}. It connects the temperature dependence of chemical potential $\mu$ to the entropy variation with fermion density $n$ at fixed $U$. {\it Second}, at fixed $T$ and $n$, the total differential $\dd n$ follows
\begin{equation}\begin{aligned}
\label{eq:TotDiffn2}
\dd n = \Big(\frac{\partial n}{\partial\mu}\Big)_{T,U}\dd\mu + \Big(\frac{\partial n}{\partial U}\Big)_{T,\mu}\dd U = 0,
\end{aligned}\end{equation}
and it leads to the second Maxwell relation
\begin{equation}\begin{aligned}
\label{eq:Maxwell001}
\Big(\frac{\partial \mu}{\partial U}\Big)_{T,n}
= -\frac{\big(\frac{\partial n}{\partial U}\big)_{T,\mu}}
{\big(\frac{\partial n}{\partial\mu}\big)_{T,U}}
= -\Big(\frac{\partial h_{I}}{\partial n}\Big)_{T,U}
= \frac{1}{2} - \Big(\frac{\partial D}{\partial n}\Big)_{T,U},
\end{aligned}\end{equation}
where Eq.~(\ref{eq:Maxwellc}) and the equality $h_I = D - n/2$ are used in the second and third equality, respectively. {\it Third}, by combining the total differentials $\dd\eyps$ with fixed $T$ and $n$ as well as $\dd h_{I}$ with fixed $U$ and $n$, we can reach the third Maxwell relation
\begin{equation}\begin{aligned}
\label{eq:Maxwell002}
\Big(\frac{\partial \eyps}{\partial U}\Big)_{n,T}
= -\Big(\frac{\partial D}{\partial T}\Big)_{n,U}.
\end{aligned}\end{equation}
The detailed derivation is summarized in Appendix~\ref{Sec:A2Entropy}. This relation associates the entropy variation with $U$ to the temperature derivative of double occupancy $D$. For the special case of $n=1$, Eq.~(\ref{eq:Maxwell002}) becomes identical to Eq.~(\ref{eq:Maxwella}) obtained from a fixed-$\mu$ condition, and it also reduces to the half-filling form $(\partial\eyps/\partial U)_{T}=-(\partial D/\partial T)_{U}$. At arbitrary fixed filling $n$, double occupancy $D=n^2/4$ at $U=0$ holds, and according to Eq.~(\ref{eq:Maxwell002}), it leads to $(\partial D/\partial T)_{n,U=0}=0$ and subsequently results in $(\partial\eyps/\partial U)_{n,T}=0$ at $U=0$. This means that the $U=0$ point is the extremum of the $\eyps(U)$ function at fixed $n$ and $T$. Besides, based on the entropy formula in Eq.~(\ref{eq:changeUfixTn1}), we can actually evaluate the derivative $(\partial \eyps/\partial U)_{n,T}$ as 
\begin{equation}\begin{aligned}
\label{eq:ParSParU}
\Big(\frac{\partial \eyps}{\partial U}\Big)_{n,T}
&= \frac{1}{T}\Big[ \frac{\partial \widetilde{e}_{\rm in}(U)}{\partial U} - D(U) \Big]_{n,T} \\
&= \frac{1}{T}\Big[ \frac{\partial e_k(U)}{\partial U} + U\frac{\partial D(U)}{\partial U}\Big]_{n,T},
\end{aligned}\end{equation}
where $\widetilde{e}_{\rm in}(U)=e_k(U)+UD(U)$ is applied in the second equality. Equation~(\ref{eq:ParSParU}) can be used to numerically estimate the extremum of $\eyps(U)$ by solving $(\partial\eyps/\partial U)_{n,T}=0$.

In the Hubbard model, Maxwell relations provide explicit connections between different thermodynamic observables, including thermal entropy $\eyps$, double occupancy $D$, chemical potential $\mu$, and fermion density $n$. From our calculations, Eqs.~(\ref{eq:Maxwella})--(\ref{eq:Maxwellc}), associated with Eq.~(\ref{eq:ParSParMu2nd}), are the Maxwell relations for the fixed-$\mu$ simulations of the doped Hubbard model~(\ref{eq:Hamiltonian}), while the fixed-$n$ correspondences include Eqs.~(\ref{eq:Maxwell000}),~(\ref{eq:Maxwell001}), and~(\ref{eq:Maxwell002}). A clear understanding of these relations can offer valuable insights into the essential physics in the doped Hubbard model.

\subsection{Tuning \texorpdfstring{$\mu$}{} to reach a fixed fermion filling}
\label{Sec:TuneMu}

In grand canonical ensemble simulations for the Hubbard model, the requirement of fixing fermion filling $n$ to a desired value $n_{\rm target}$ is frequently encountered in various quantum many-body numerical approaches~\cite{Macridin2009,Khatami2009,Khatami2011,Ehsan2011,Lenihan2021,YuanYao2019B,Qiaoyi2026}. This is typically realized via tuning the chemical potential $\mu$ by hand, based on the observation that $n(\mu)$ versus $\mu$ is a monotonically decreasing function in the model~(\ref{eq:Hamiltonian}). Here, we introduce a highly efficient procedure for tuning $\mu$ based on the charge compressibility, following the idea discussed in Ref.~\cite{Miles2022}.

We start with a relatively wide range of $\mu$, denoted as $[\mu_{i},\mu_{f}]$, which contains the final solution $\mu_{\rm target}$ satisfying $n(\mu_{\rm target})=n_{\rm target}$. First, we apply the standard bisection or linear interpolation method to update $(\mu_{i},\mu_{f})$ and to narrow down the correct range containing $\mu_{\rm target}$. For instance, given that $n(\mu_i)>n_{\rm target}$ and $n(\mu_f)<n_{\rm target}$, we generate $\mu_{\rm new}=(\mu_i+\mu_f)/2$ via bisection (or an analogous $\mu_{\text{new}}$ via linear interpolation) and compute $n(\mu_{\rm new})$, and then set $\mu_i=\mu_{\rm new}$ if $n(\mu_{\rm new})>n_{\rm target}$ or otherwise set $\mu_f = \mu_{\text{new}}$. Second, once the difference $(\mu_f-\mu_i)$ is reduced to a threshold (e.g., $\sim10^{-2}t$), the desired fermion density $n(\mu_{\rm target})=n_{\rm target}$ can be approximated by its first-order Taylor expansion around $\mu_0=\mu_{\rm new}$ from the last step as
\begin{equation}\begin{aligned}
\label{eq:Taylorn}
n_{\rm target} \simeq n(\mu_0) + (\mu_{\rm target}-\mu_0) \frac{\partial n(\mu)}{\partial\mu}\Big|_{\mu=\mu_0},
\end{aligned}\end{equation}
in which $\partial n(\mu)/\partial\mu$ is related to the charge compressibility $\kappa$ that can be numerically computed via~\cite{Song2025B}
\begin{equation}\begin{aligned}
\label{eq:Chie}
\kappa 
= - \frac{\partial n(\mu)}{\partial\mu} 
= \frac{\beta}{N_s}\sum_{\mathbf{ij}}\big(\langle \hat{n}_{\mathbf{i}} \hat{n}_{\mathbf{j}} \rangle - \langle \hat{n}_{\mathbf{i}} \rangle\langle \hat{n}_{\mathbf{j}} \rangle\big).
\end{aligned}\end{equation}
The calculation of $\kappa$ only involves the density-density correlation function, which is straightforward in numerical methods. From Eq.~(\ref{eq:Taylorn}), we obtain an approximate solution $\widetilde{\mu}_{\rm target}$ for the true $\mu_{\rm target}$ as
\begin{equation}\begin{aligned}
\label{eq:Recursive}
\widetilde{\mu}_{\rm target} = \mu_0 +\frac{n(\mu_0)-n_{\rm target}}{\kappa(\mu_0)}.
\end{aligned}\end{equation}
Then we set $\mu_0=\widetilde{\mu}_{\rm target}$, compute $n(\mu_0)$, apply Eq.~(\ref{eq:Recursive}) to get a new $\widetilde{\mu}_{\rm target}$, and repeat the process until convergence. In statistical methods such as AFQMC, $n(\mu_0)$ from the simulation has average and error bar, denoted as $\bar{n}_0$ and $\sigma_{n_0}$. We typically use $|\bar{n}_0-n_{\rm target}|\le\sigma_{n_0}$ as the convergence criterion. The combination of the first and second steps as discussed above enables the precise determination of $\mu_{\rm target}$. 

In practical AFQMC simulations, we find that the above procedure can converge for most situations within quite a limited number of iterations, such as three in the first step and two in the second. 
Moreover, since $\mu_{\rm target}$ typically varies only slightly with changes in adjacent model parameters or linear system sizes, a more accurate guess for the initial range $[\mu_i, \mu_f]$ can often be made in subsequent simulations, thereby accelerating convergence. More broadly, the above self-consistent procedure for determining $\mu_{\rm target}$ can be applied to other quantum many-body methods, provided that $n(\mu)$ and $\kappa(\mu)$ can be computed efficiently and accurately.

\section{An alternative derivation for the entropy evaluation formulas}
\label{Sec:Hellmann}

In the previous section, we derived the entropy evaluation formulas from the total differential of the grand potential. Here, we provide an alternative derivation based on the connection between the grand potential and the partition function, combined with the finite-temperature Hellmann-Feynman theorem. 

In thermodynamics and statistical physics, the grand potential density $\gpps$ is connected to the partition function $Z=\text{Tr}(e^{-\beta\hat{H}})$ via $\gpps=-N_s^{-1}T\ln Z$. The ensemble average of an observable $\hat{O}$ reads $\langle \hat{O} \rangle=Z^{-1}{\Tr}(e^{-\beta\hat{H}}\hat{O})$. From the model~(\ref{eq:Hamiltonian}), one can readily obtain the following derivative
\begin{equation}\begin{aligned}
\label{eq:GrandTdiff}
\frac{\dd (\gpps/T)}{\dd T} = -\frac{e_{\mathrm{tot}}(T)}{T^2} + \frac{n}{T} \frac{\dd \mu(T)}{\dd T}.
\end{aligned}\end{equation}
Note that the chemical potential $\mu$ can be a function of temperature $T$ in the fixed-$n$ calculations. This equality enables the direct evaluation of $\gpps/T$ via integration and subsequently the thermal entropy along the $T$ axis. For other paths in the parameter space, such as those along the $\mu$ or $U$ axes, we rely on the finite-temperature Hellmann-Feynman theorem
\begin{equation}\begin{aligned}
\label{eq:HFfiniteT}
\frac{\partial \gpps}{\partial \alpha} = \Big\langle \frac{\partial \hat{H}}{\partial \alpha} \Big\rangle,
\end{aligned}\end{equation}
which can be proved using the Lehmann representation (see Appendix A in Ref.~\cite{Song2025B}). By setting $\alpha=\mu$ and $\alpha=U$, we obtain explicit expressions for $\partial\gpps/\partial\mu$ and $\partial\gpps/\partial U$. Integrating these expressions gives $\gpps(\mu)$ and $\gpps(U)$, which in turn allow the thermal entropy to be evaluated. In the following, we combine Eqs.~(\ref{eq:GrandTdiff}),~(\ref{eq:HFfiniteT}), and~(\ref{eq:GPdensity}) to independently derive the entropy formulas obtained in Sec.~\ref{Sec:Ensemble} for four different paths.

For varying $T$ with fixed $U$ and $n$, we can take an integration over temperature from $T$ to $\infty$ for both sides of Eq.~(\ref{eq:GrandTdiff}). In this calculation, the limit $[\gpps(T)/T]_{T=\infty}$ needs to be carefully handled. From Eq.~(\ref{eq:GPdensity}), $\gpps(T)/T=e_{\rm in}/T + \mu n/T-\eyps$. As $T\to\infty$, $\mu$ should diverge to reach the fixed filling $n$ (see Appendix~\ref{Sec:A2Entropy}), while $e_{\rm in}$ remains finite. This leads to $[\gpps(T)/T]_{T=\infty}=(n\mu/T)_{T=\infty}-\eyps_n(\infty)$, where $\eyps_n(\infty) = 2\ln2 - n\ln n - (2-n)\ln(2-n)$ is the fixed-$n$ entropy density at $T=\infty$ (see Appendix~\ref{Sec:A2Entropy}). Thus, after the integration, we arrive at
\begin{equation}\begin{aligned}
\label{eq:OmegaT1}
\frac{\gpps(T)}{T}=
&-\eyps_n(\infty) +n\frac{\mu(T)}{T}\Big{|}_{T=\infty} \\
& +\int_{T}^{\infty} \frac{e_{\mathrm{tot}}(T')}{T^{'2}} \dd T' - n \int_{T}^{\infty} \frac{1}{T'} \dd \mu(T').
\end{aligned}\end{equation}
Then applying the relation $e_{\rm tot}=e_{\rm in}+\mu n$ and performing the integration by parts for both integrals, the above equality can be simplified as
\begin{equation}\begin{aligned}
\label{eq:GrandCompt1}
\frac{\gpps(T)}{T} = -\eyps_n(\infty) + n\frac{\mu(T)}{T} + \int_{T}^{\infty} \frac{e_{\mathrm{in}}(T')}{T^{'2}} \dd T'.
\end{aligned}\end{equation}
Then combining Eqs.~(\ref{eq:GPdensity}) and~(\ref{eq:GrandCompt1}), the entropy formula in Eq.~(\ref{eq:fixedn}) can be exactly reproduced. 

For varying $T$ with fixed $U$ and $\mu$, the derivative $\dd\mu/\dd T$ in Eq.~(\ref{eq:GrandTdiff}) vanishes. For this case, the total energy $e_{\rm tot}$ is always finite, leading to $[\gpps(T)/T]_{T=\infty}=-\eyps_{\mu}(\infty)$ with $\eyps_{\mu}(\infty)=\ln 4$ as the fixed-$\mu$ entropy density at $T=\infty$. Then the integration over temperature for Eq.~(\ref{eq:GrandTdiff}) yields
\begin{equation}\begin{aligned}
\label{eq:GrandCompt2}
\frac{\gpps(T)}{T} = -\eyps_{\mu}(\infty) + \int_{T}^{\infty} \frac{e_{\mathrm{tot}}(T')}{T^{'2}} \dd T',
\end{aligned}\end{equation}
which evidently recovers the formulas in Eqs.~(\ref{eq:fixedmT_rough}) and~(\ref{eq:fixedmT}) when combining $\gpps(T)/T=e_{\rm tot}/T-\eyps(T)$ from Eq.~(\ref{eq:GPdensity}).

For varying $\mu$ with fixed $U$ and $T$, we apply Eq.~(\ref{eq:HFfiniteT}) and choose $\alpha=\mu$, which gives
\begin{equation}\begin{aligned}
\frac{\partial \gpps(\mu)}{\partial \mu} = \Big\langle \frac{\partial\hat{H}}{\partial \mu}\Big\rangle=n. 
\end{aligned}\end{equation}
Integrating over $\mu$ for this equality gives
\begin{equation}\begin{aligned}
\label{eq:GrandCompt3}
\gpps(\mu) = \gpps(0) + \int_0^\mu n(\mu')\dd \mu',
\end{aligned}\end{equation}
which clearly reproduces the formula in Eq.~(\ref{eq:changem}) considering $\gpps(\mu)=e_{\rm tot}-T\eyps(\mu)$.

For varying $U$ with fixed $T$ and $n$, we choose $\alpha=U$ in Eq.~(\ref{eq:HFfiniteT}), and reach the Hellmann-Feynman theorem
\begin{equation}\begin{aligned}
\frac{\partial \gpps(U)}{\partial U} = \Big\langle \frac{\partial\hat{H}}{\partial U}\Big\rangle= h_I + n\frac{\partial\mu}{\partial U}. 
\end{aligned}\end{equation}
The integration over $U$ for this equality yields
\begin{equation}\begin{aligned}
\label{eq:GrandCompt4}
\gpps(U) = \gpps(0) + \int_0^{U} h_I(U')\dd U'  + n[\mu(U)-\mu(0)],
\end{aligned}\end{equation}
which recovers the formula in Eq.~(\ref{eq:changeUIni1}) by combining $\gpps(U)=e_{\rm in} + \mu(U) n - T\eyps(U)$. 

In addition to reproducing the entropy formulas, the expressions in Eqs.~(\ref{eq:GrandCompt1}),~(\ref{eq:GrandCompt2}),~(\ref{eq:GrandCompt3}), and~(\ref{eq:GrandCompt4}) can also be used to directly compute the grand potential density $\gpps$. This quantity plays a central role in the thermodynamics of quantum systems, as many commonly used observables can be obtained from its derivatives. In Ref.~\cite{Rampon2025}, $\gpps$ is employed as a key quantity to identify possible incommensurate orders in the 3D doped Hubbard model. Therefore the reliable computation of $\gpps$ in the doped Hubbard model can provide additional insights into its finite-temperature properties.

\begin{figure}
\centering
\includegraphics[width=0.965\columnwidth]{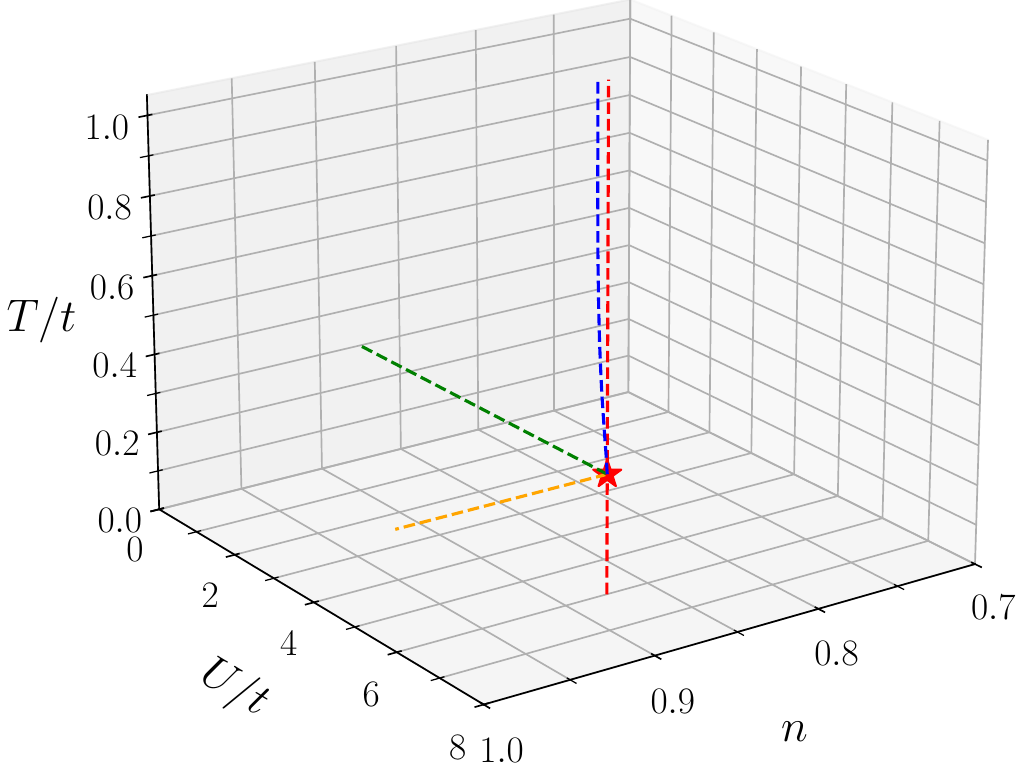}
\caption{Illustration of four calculation schemes for the thermal entropy along different paths in the parameter space of the doped Hubbard model, consisting of interaction strength $U/t$, temperature $T/t$, and fermion filling $n$. The red dashed line shows the varying-$T$ path with fixed $U/t=6$ and $n=0.875$. The blue dashed line plots the varying-$T$ path with fixed $U/t=6$ and $\mu=\mu_0$. The orange dashed line represents the varying-$\mu$ path with fixed $U/t=6$ and $T/t=0.3$. The green dashed line denotes the varying-$U$ path with fixed $T/t=0.3$ and $n=0.875$. The value $\mu_0\simeq 1.23656$ is chosen such that $n(\mu_0)=0.875$ at $(U/t=6,T/t=0.3)$ for the $L=4$ system in the 3D Hubbard model. This reference point is marked by the red star.}
\label{fig:Fig01Paths}
\end{figure}

\section{Numerical results}
\label{Sec:Numeric}

Applying the calculation formulas derived in Sec.~\ref{Sec:Ensemble}, we present numerical results for the thermal entropy obtained from finite-temperature AFQMC simulations in this section. First, we test the entropy evaluation schemes as well as the Maxwell relations in the 3D doped Hubbard model using a small system with $L=4$ for simplicity. The test results are summarized in Sec.~\ref{Sec:Testin3D}. Second, we perform large-scale simulations for physically relevant parameters in the {2D} doped Hubbard model and benchmark the resulting entropy against that reported in a previous DiagMC study. The corresponding results and discussions are presented in Sec.~\ref{Sec:PhysicsIn{2D}}. The parameter grids and Monte Carlo sampling details used in these calculations are summarized in Appendix~\ref{Sec:A3Simulation}.

In Fig.~\ref{fig:Fig01Paths}, we illustrate the calculation schemes for the thermal entropy in the doped Hubbard model along four paths in its parameter space, corresponding to the formulas achieved in Sec.~\ref{Sec:Ensemble}. For the numerical tests, we consider the 3D Hubbard model~(\ref{eq:Hamiltonian}) on the simple cubic lattice with $L=4$, and all four paths cross at $(U/t=6$, $T/t=0.3$, $n=0.875)$. For the varying-$U$ path, we perform the $U$-sweep calculations from $U/t=-6$ to $U/t=+6$. For the other three paths, we carry out the calculations for $U/t=+6$, $+4$, $+2$, $-2$, $-4$, $-6$. All the fixed-$n$ simulations are realized via tuning the chemical potential $\mu$ to reach the target filling $n=0.875$, applying the procedure discussed in Sec.~\ref{Sec:TuneMu}. For the physically relevant results in the {2D} doped Hubbard model, we concentrate on the varying-$\mu$ path with fixed $U/t$ and $T/t$, and perform simulations up to $L=16$, which clearly show convergence toward the thermodynamic limit. 

\begin{figure}[t]
\centering
\includegraphics[width=1.00\columnwidth]{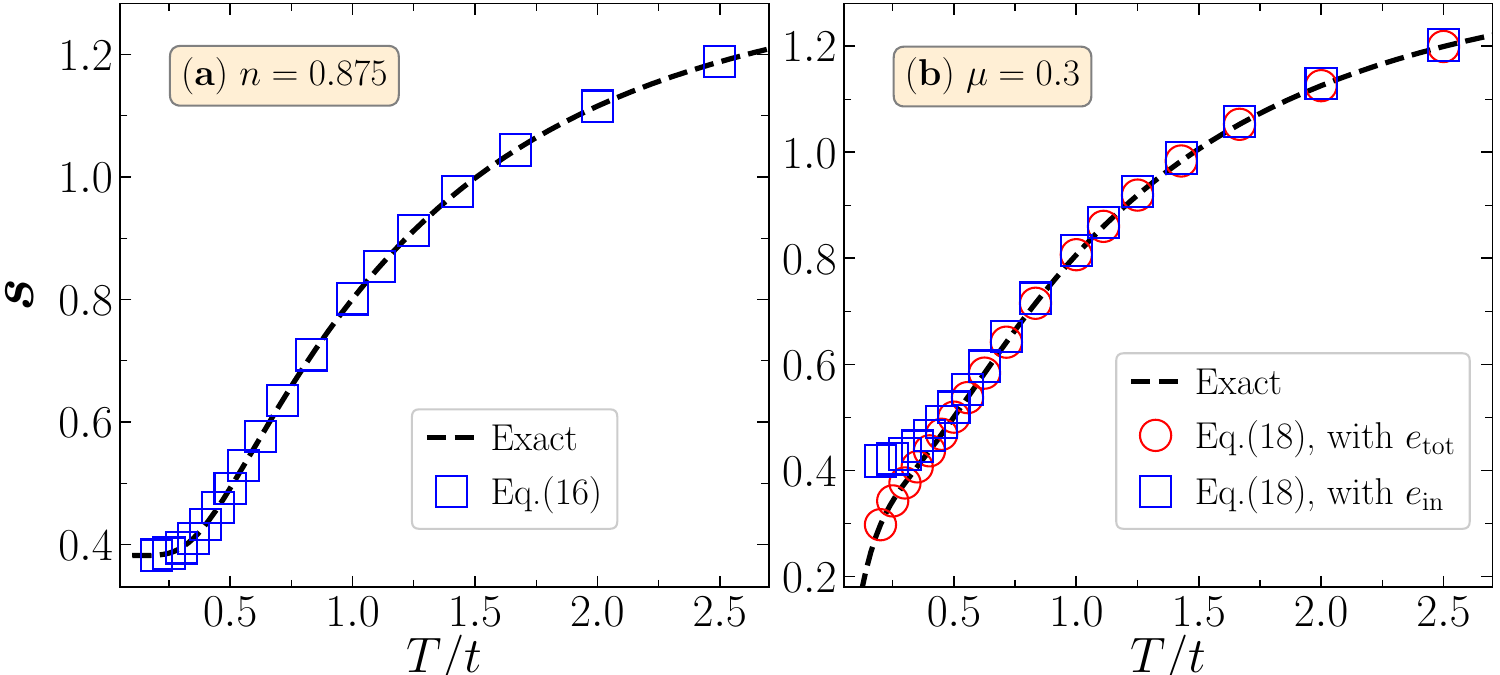}
\caption{Entropy density $\eyps$ as a function of $T/t$ for the 3D Hubbard model with $U=0$ from $L=4$ system. (a) plots the results (blue squares) for fixed $n=0.875$ evaluated from Eq.~(\ref{eq:fixednT}), while (b) shows the fixed-$\mu$ results (red circles) with $\mu/t=0.3$ obtained from Eq.~(\ref{eq:fixedmT}). In both panels, the black dashed line denotes the exact results computed via Eq.~(\ref{eq:U0Exact}). In (b), the blue squares are obtained from Eq.~(\ref{eq:fixedmT}) using the incorrect energy $e_{\rm in}$ rather than $e_{\rm tot}$, yielding intentionally erroneous results.}
\label{fig:Fig02U0Entropy}
\end{figure}

\subsection{Test results in the 3D doped Hubbard model}
\label{Sec:Testin3D}

We begin with testing the varying-$T$ calculation schemes for entropy density $\eyps$ in the noninteracting system. At $U=0$, the exact results of $\eyps$ can be computed via Eq.~(\ref{eq:U0Exact}) for either fixed $n$ or fixed $\mu$. In Fig.~\ref{fig:Fig02U0Entropy}, we show the comparisons between the exact results and those from the numerical calculations using Eqs.~(\ref{eq:fixednT}) and~(\ref{eq:fixedmT}). For both the fixed-$n$ and fixed-$\mu$ situations, the perfect match is clear, validating the evaluation formulas for $\eyps$. For the fixed-$\mu$ calculation, we also include results intentionally miscalculated using Eq.~(\ref{eq:fixedmT}) with the energy $e_{\rm in}$ instead of the correct $e_{\rm tot}$, shown as blue squares in Fig.~\ref{fig:Fig02U0Entropy}(b). These conceptually incorrect results almost coincide with the correct ones at high temperatures but clearly deviate for $T/t \le 0.5$. This can be understood from the relation $e_{\rm tot}=e_{\rm in}+\mu n$, combined with the fact that for fixed $\mu$, the system gradually approaches half filling with $n=1$ in the atomic limit. In the high-$T$ regime, this renders the additional energy $\mu n$ to be nearly a constant, thereby eliminating its contribution in Eq.~(\ref{eq:fixedmT}). In the low-$T$ regime, by contrast, $n$ varies significantly with $T$, and the $\mu n$ term in Eq.~(\ref{eq:fixedmT}) contributes substantially, leading to the deviation observed in Fig.~\ref{fig:Fig02U0Entropy}(b).

\begin{figure}[t]
\centering
\includegraphics[width=0.953\columnwidth]{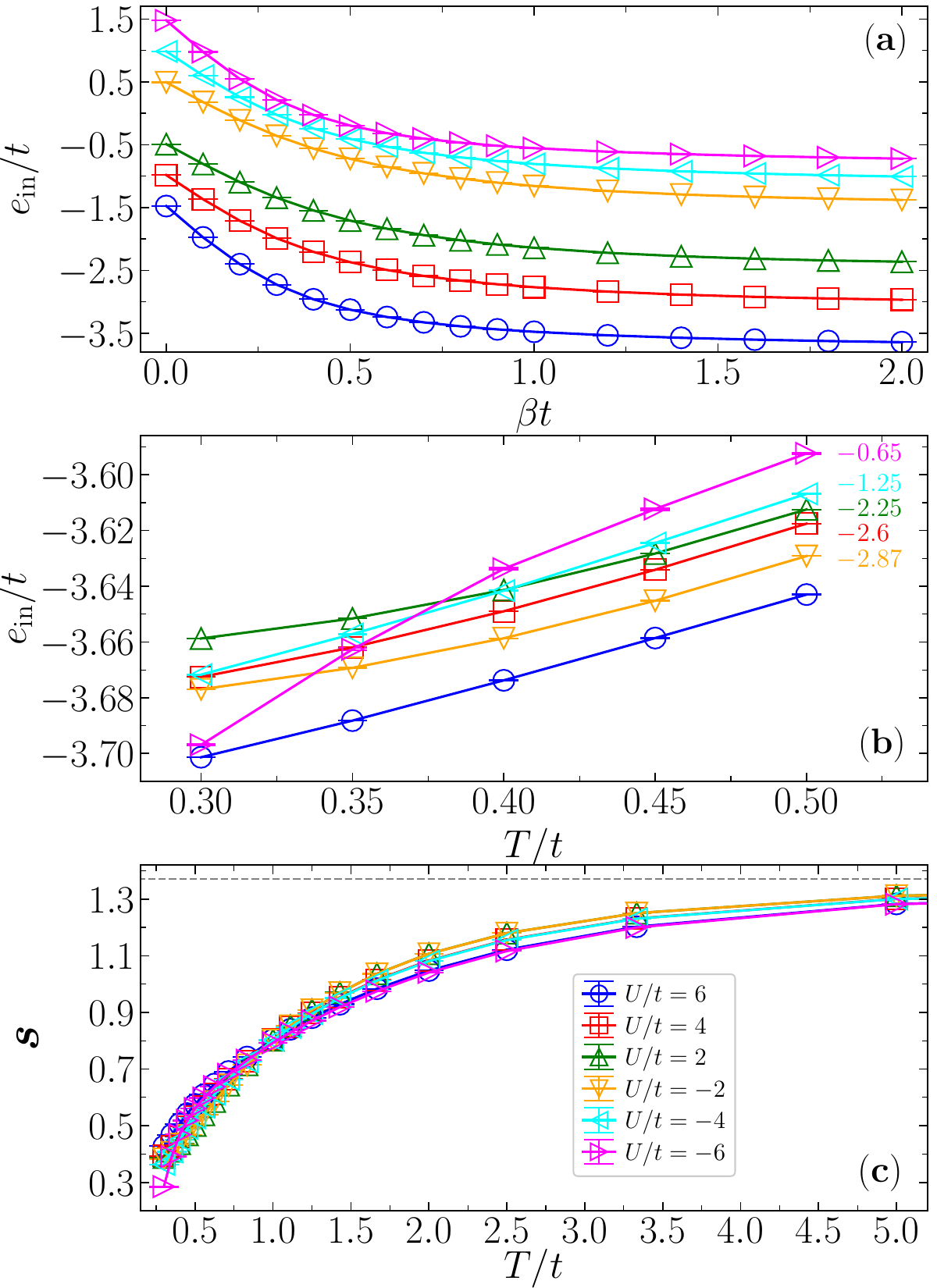}
\caption{Test results for the varying-$T$ path with fixed $U/t$ and $n=0.875$. (a) and (b) plot the energy density $e_{\mathrm{in}}/t$ vs inverse temperature $\beta t$ within $0\le \beta t\le 2.0$ and vs temperature $T/t$ within $0.30\le T/t\le 0.50$, respectively. The $e_{\rm in}/t$ data in (b) have been vertically offset (except for $U/t=6$) to fit into the plot, and the shifts are marked with the same colors as the data. (c) shows the entropy density $\eyps$ as a function of $T/t$, computed via Eq.~(\ref{eq:fixednT}) with $T_0/t=0.5$ and $\beta_0 t=2.0$. The calculations are performed on an $L=4$ system for the 3D Hubbard model with $U/t=+6$, $+4$, $+2$, $-2$, $-4$, $-6$. The black dashed line in (c) plots $\eyps_n(\infty) = \ln 4 - n \ln n - (2 - n) \ln (2 - n)$ as the entropy result at $T/t=\infty$.}
\label{fig:Fig03EntropyPath1}
\end{figure}

We then turn to the entropy evaluations in the interacting system. The test results are demonstrated in the following according to the four calculation schemes outlined in Sec.~\ref{Sec:Ensemble}. In the AFQMC simulations, we mostly adopt $\Delta\tau t=0.04$, and specifically take $\Delta\tau t=0.015$ for the varying-$U$ simulations due to the significant Trotter error in double occupancy (see Appendix~\ref{Sec:A1douocc}).

For the varying-$T$ path with fixed $U/t$ and $n=0.875$, we plot the test results in Fig.~\ref{fig:Fig03EntropyPath1}, including energy density $e_{\rm in}/t$ and entropy density $\eyps$ computed via Eq.~(\ref{eq:fixednT}). The results of $e_{\rm in}/t$ in Figs.~\ref{fig:Fig03EntropyPath1}(a) and~\ref{fig:Fig03EntropyPath1}(b) are presented according to the intermediate $T_0=1/\beta_0$ in Eq.~(\ref{eq:fixednT}), which is chosen as $T_0/t=0.5$ (and $\beta_0=2.0$). As expected, $e_{\rm in}/t$ is rather smooth in both ranges of $0\le\beta\le\beta_0$ and $T\le T_0$, thereby allowing us to accurately evaluate the integrals in Eq.~(\ref{eq:fixednT}) with a few data points in each range. The double occupancy $D=n^2/4$ at $T/t=\infty$ and the associated energy $e_{\rm in}=U(D-n/2)$, are directly used at $\beta t=0$ in Fig.~\ref{fig:Fig03EntropyPath1}(a). The resulting $\eyps$ in Fig.~\ref{fig:Fig03EntropyPath1}(c) evidently shows the convergence to $\eyps_n(T=\infty)$ as $T/t$ increases. Interestingly, we find that the numerical results of $\eyps(+U)$ and $\eyps(-U)$ almost coincide in the high-$T$ regime (such as $T/t\ge 1.0$). This behavior can be understood from the atomic limit ($\beta|U|\ll 1$), which dominates the high-$T$ physics of the Hubbard model~\cite{Song2025B}. In this regime, the kinetic energy $e_k$ is small and largely insensitive to the sign of $U$, and thus $e_{\rm in}$ is governed by the interaction energy $U(D-n/2)$. As $\beta t$ increases from zero, the reduction of $e_{\rm in}/t$ arises from the opposite temperature dependence of the double occupancy: it is suppressed upon cooling for $U>0$, but enhanced for $U<0$. Moreover, the atomic limit result gives (see Appendix~\ref{Sec:A2Entropy})
\begin{equation}\begin{aligned}
\label{eq:AtomicD}
D(T,U) = \frac{n^2}{4} - \frac{n^2(2-n)^2}{16}\beta U + \mathcal{O}[(\beta U)^2].
\end{aligned}\end{equation}
This equality together with the above analysis indicates that the energy difference between $e_{\rm in}(-U)$ and $e_{\rm in}(+U)$ in the high-$T$ regime can be approximated as
\begin{equation}\begin{aligned}
\label{eq:EDiff}
e_{\rm in}(-U)-e_{\rm in}(+U) 
& \simeq -U \big[D(T,-U)+D(T,+U) - n\big] \\
& = +U(n - n^2/2),
\end{aligned}\end{equation}
which is a $T$-independent constant that only depends on the filling $n$. This qualitatively explains the behavior of $e_{\rm in}/t$ results in Fig.~\ref{fig:Fig03EntropyPath1}(a), where the curves for $e_{\rm in}(-U)$ and $e_{\rm in}(+U)$ are approximately related by a constant shift. More quantitatively, here for $n=0.875$ in our calculations, the energy difference predicted by Eq.~(\ref{eq:EDiff}) is $\sim 0.492U$, which agrees well with the shifts observed for all three pairs of $(-U,+U)$. Regarding the $\eyps$ results in Fig.~\ref{fig:Fig03EntropyPath1}(c), since the contribution from a $T$-independent constant in $e_{\rm in}$ cancels out in Eq.~(\ref{eq:fixednT}), $\eyps(+U)$ and $\eyps(-U)$ naturally become almost identical in the high-$T$ regime. In the low-$T$ regime, the atomic-limit approximation and accordingly Eqs.~(\ref{eq:AtomicD}) and~(\ref{eq:EDiff}) are instead no longer valid, which leads to the deviation between $\eyps(+U)$ and $\eyps(-U)$ as shown in Fig.~\ref{fig:Fig03EntropyPath1}(c). This deviation becomes more pronounced for larger $|U|$, since the atomic-limit condition $\beta |U|\ll 1$ breaks down earlier as $T/t$ decreases. Besides, $\eyps$ always decreases monotonically upon cooling, as anticipated.

\begin{figure}[t]
\centering
\includegraphics[width=0.955\columnwidth]{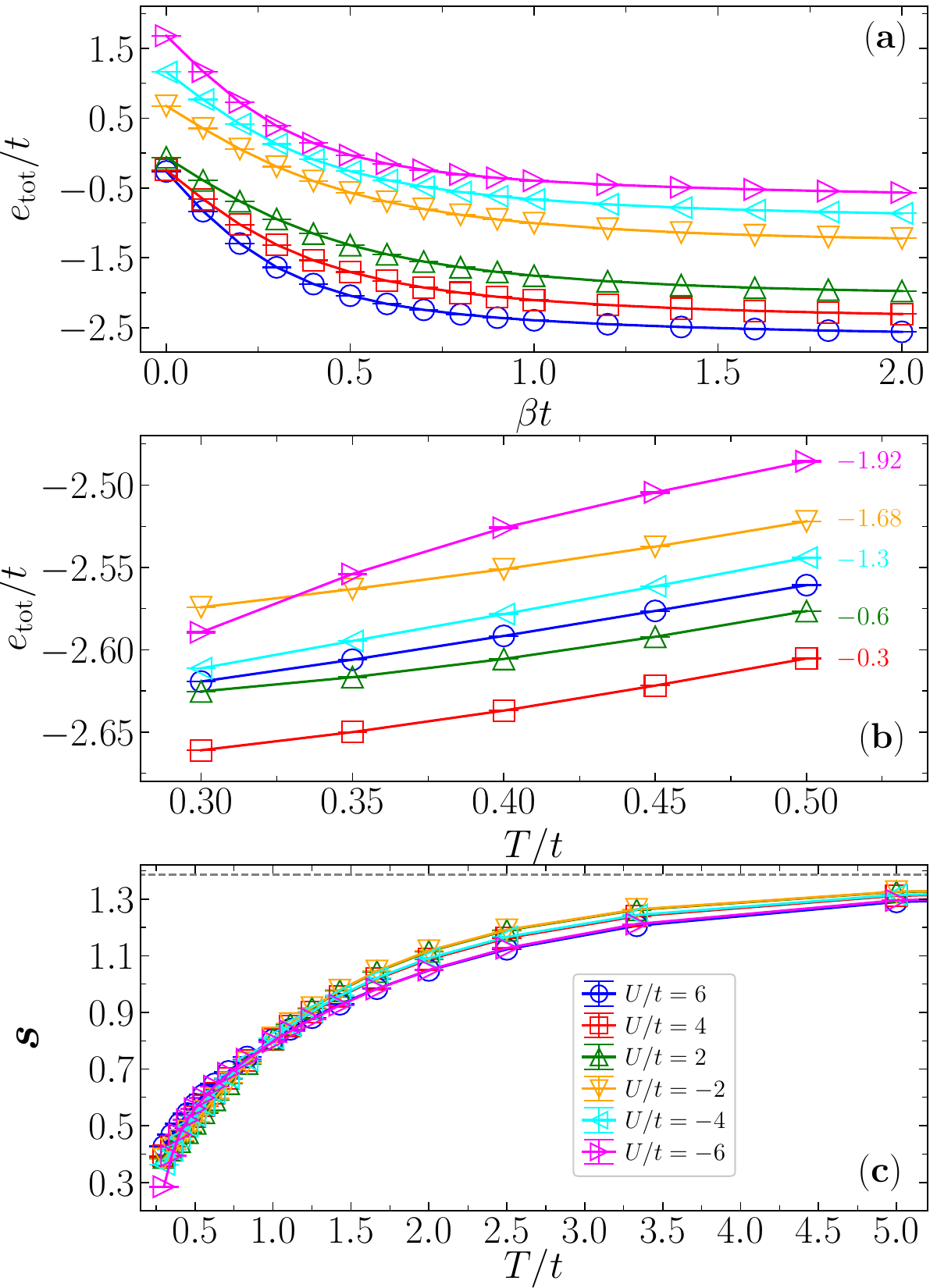}
\caption{Test results for the varying-$T$ path with fixed $U/t$ and $\mu/t$. (a) and (b) plot the energy density $e_{\mathrm{tot}}/t$ vs inverse temperature $\beta t$ within $0\le \beta t\le 2.0$ and vs temperature $T/t$ within $0.30\le T/t\le 0.50$, respectively. The $e_{\rm tot}/t$ data in (b) have been vertically offset (except for $U/t=6$) to fit into the plot, and the shifts are marked with the same colors as the data. (c) shows the entropy density $\eyps$ as a function of $T/t$, computed via Eq.~(\ref{eq:fixedmT}) with $T_0/t=0.5$ and $\beta_0 t=2.0$. The calculations are performed on an $L=4$ system for the 3D Hubbard model with $U/t=+6$, $+4$, $+2$, $-2$, $-4$, $-6$, at fixed chemical potential $\mu = \mu_0$, chosen such that $n(\mu_0) = 0.875$ at $T/t = 0.3$ for all values of $U/t$. The black dashed line in (c) plots $\eyps_{\mu}(\infty) = \ln 4$ as the entropy result at $T/t=\infty$.}
\label{fig:Fig04EntropyPath2}
\end{figure}

For the varying-$T$ path with fixed $U/t$ and $\mu/t$, we plot the test results in Fig.~\ref{fig:Fig04EntropyPath2}, in a manner similar to Fig.~\ref{fig:Fig03EntropyPath1}. The fixed $\mu$ is chosen as $\mu=\mu_0$ such that $n(\mu_0) = 0.875$ at $T/t = 0.3$ for all values of $U/t$. Apparently, $\mu_0$ differs for different $U/t$. For example, our AFQMC simulations identify $\mu_0/t\simeq 0.17997$ and $1.23656$ for $U/t=-6$ and $+6$, respectively. Here the entropy density $\eyps$ is computed using Eq.~(\ref{eq:fixedmT}) with $T_0/t=0.5$ and $\beta_0 t=2.0$. The results of energy density $e_{\rm tot}/t$ are also smooth in both ranges of $0\le\beta\le\beta_0$ and $T\le T_0$. Similar to Fig.~\ref{fig:Fig03EntropyPath1}, we find that the curves for $e_{\rm tot}(-U)$ and $e_{\rm tot}(+U)$ are also approximately related by a constant shift in the high-$T$ regime, as shown in Fig.~\ref{fig:Fig04EntropyPath2}(a). This can again be explained from the atomic-limit approximation, which renders $n=1$ for finite $\mu$ under the $\beta|U|\ll1$ condition. Thus Eq.~(\ref{eq:EDiff}) is modified to
\begin{equation}\begin{aligned}
\label{eq:EtotDiff}
e_{\rm tot}(-U)-e_{\rm tot}(+U) \simeq U/2 + \mu_0(-U) - \mu_0(+U),
\end{aligned}\end{equation}
considering $e_{\rm tot}=e_{\rm in} + \mu n$. For $U/t=6$, this energy difference is $\sim 1.943t$, which quantitatively agrees with the numerical results in Fig.~\ref{fig:Fig04EntropyPath2}(a). For the other two pairs of $(-U,+U)$, similar agreement is also confirmed in our calculations. Then this nearly constant difference between $e_{\rm tot}(-U)$ and $e_{\rm tot}(+U)$ also leads to the almost merged results of $\eyps(-U)$ and $\eyps(+U)$ in the high-$T$ regime, as shown in Fig.~\ref{fig:Fig04EntropyPath2}(c). The underlying reason is similar to that in the fixed-$n$ case, as the contribution from a $T$-independent constant in $e_{\rm tot}$ cancels out in Eq.~(\ref{eq:fixedmT}). Turning to the lower-$T$ regime, the above behavior persists in $e_{\rm tot}(\pm U)$ and $\eyps(\pm U)$ for $U/t=\pm2$ and $\pm4$, as shown in Fig.~\ref{fig:Fig04EntropyPath2}(b). By contrast, the prominent deviation in $\eyps$ for $U/t=\pm 6$ originates from the obviously different curves of $e_{\rm tot}$.

\begin{figure}[t]
\centering
\includegraphics[width=0.95\columnwidth]{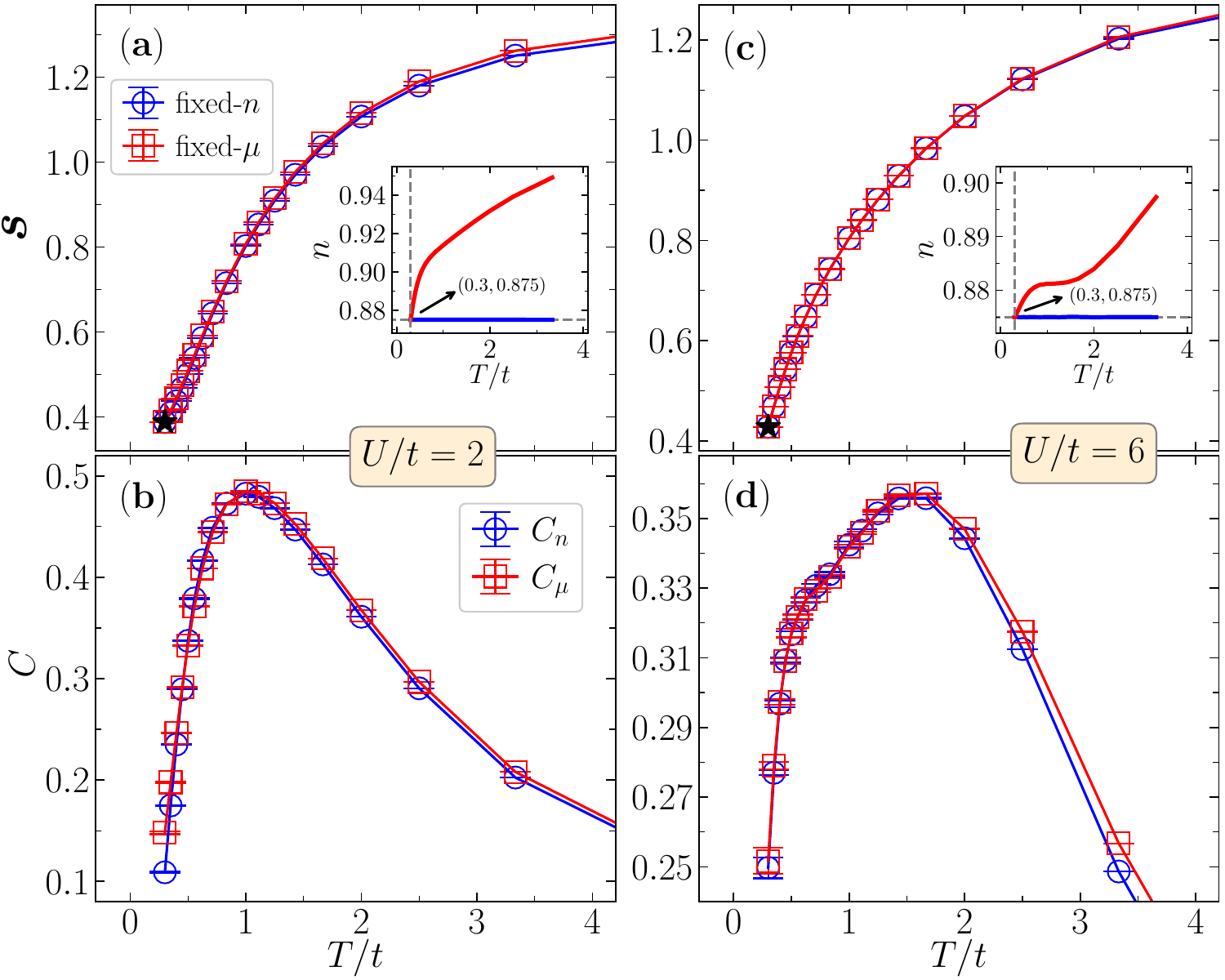}
\caption{Comparison between the fixed-$n$ and fixed-$\mu$ calculations. (a) and (b) show the results of entropy density $\eyps$ and two types of specific heat [$C_n$ and $C_{\mu}$, see Eq.~(\ref{eq:SpecHeat})] vs $T/t$, for $U/t=2$. In the fixed-$n$ path, the filling is maintained at $n=0.875$, while for the fixed-$\mu$ path, the chemical potential is set to $\mu=\mu_0$, chosen such that $n(\mu_0)=0.875$ at $T/t=0.3$. The inset in (a) plots the fermion density as a function of $T/t$ for both cases. (c) and (d) similarly show the results of $\eyps$, and $C_n$ and $C_{\mu}$, for $U/t=6$. The special point $(T/t=0.3$, $n=0.875)$ is marked by the black star in (a) and (c).}
\label{fig:Fig05sCnCmu}
\end{figure}

Based on the results in Figs.~\ref{fig:Fig03EntropyPath1} and~\ref{fig:Fig04EntropyPath2}, we further compare the fixed-$n$ and fixed-$\mu$ calculations with varying temperature for $U/t=2$ and $6$ in Fig.~\ref{fig:Fig05sCnCmu}. For each $U/t$, we present the comparison results of entropy density $\eyps$ and ($C_n,C_{\mu}$) [computed via Eq.~(\ref{eq:SpecHeat})] versus $T/t$. For the fixed-$\mu$ path, we set $\mu$ to $\mu=\mu_0$, which is chosen such that $n(\mu_0)=0.875$ at $T/t=0.3$, yielding $\mu_0/t\simeq0.43812$ and $1.23656$ for $U/t=2$ and $6$, respectively. The fixed-$n$ path maintains $n=0.875$ for both values of $U/t$. The entropy is accordingly evaluated using Eqs.~(\ref{eq:fixednT}) and~(\ref{eq:fixedmT}) in fixed-$n$ and fixed-$\mu$ calculations. We observe that the deviation in $\eyps$ between these two cases is most pronounced in the high-$T$ regime. This is reflected by their results at $T=\infty$, i.e., $\eyps_{n}(\infty)\simeq\ln 4-0.016$ for $n=0.875$ versus $\eyps_{\mu}(\infty)=\ln 4$. As $T/t$ decreases, the deviation diminishes and finally vanishes at $T/t=0.3$ [black star in Figs.~\ref{fig:Fig05sCnCmu}(a) and~\ref{fig:Fig05sCnCmu}(c)]. The specific heats $C_n$ and $C_{\mu}$ show a similar trend, with deviations at high $T$ gradually disappearing with decreasing $T/t$. The difference between $C_n$ and $C_{\mu}$ mainly comes from $e_{\rm tot}-e_{\rm in}=\mu n$ in fixed-$\mu$ calculations. Physically, we find that both $C_n$ and $C_{\mu}$ exhibit peak structures, which is the characteristic feature of the specific heat in both {2D} and 3D Hubbard models~\cite{Lu2026,Song2025B,Thereza2001}. For $U/t=2$, the peaks in $C_n$ and $C_{\mu}$ correspond to the high-$T$ peak originating from the noninteracting case around $T/t\sim 1$, while the absence of a low-$T$ peak is likely due to insufficiently low temperature or small system size. For $U/t=6$, both the prominent high-$T$ charge peak around $T/t\sim 1.5$, dominated by the atomic-limit behavior, and a weak shoulder around $T/t\sim 0.5$, serving as a precursor of the low-$T$ spin peak, can be observed even within $L=4$ system.

\begin{figure}[t]
\centering
\includegraphics[width=0.952\columnwidth]{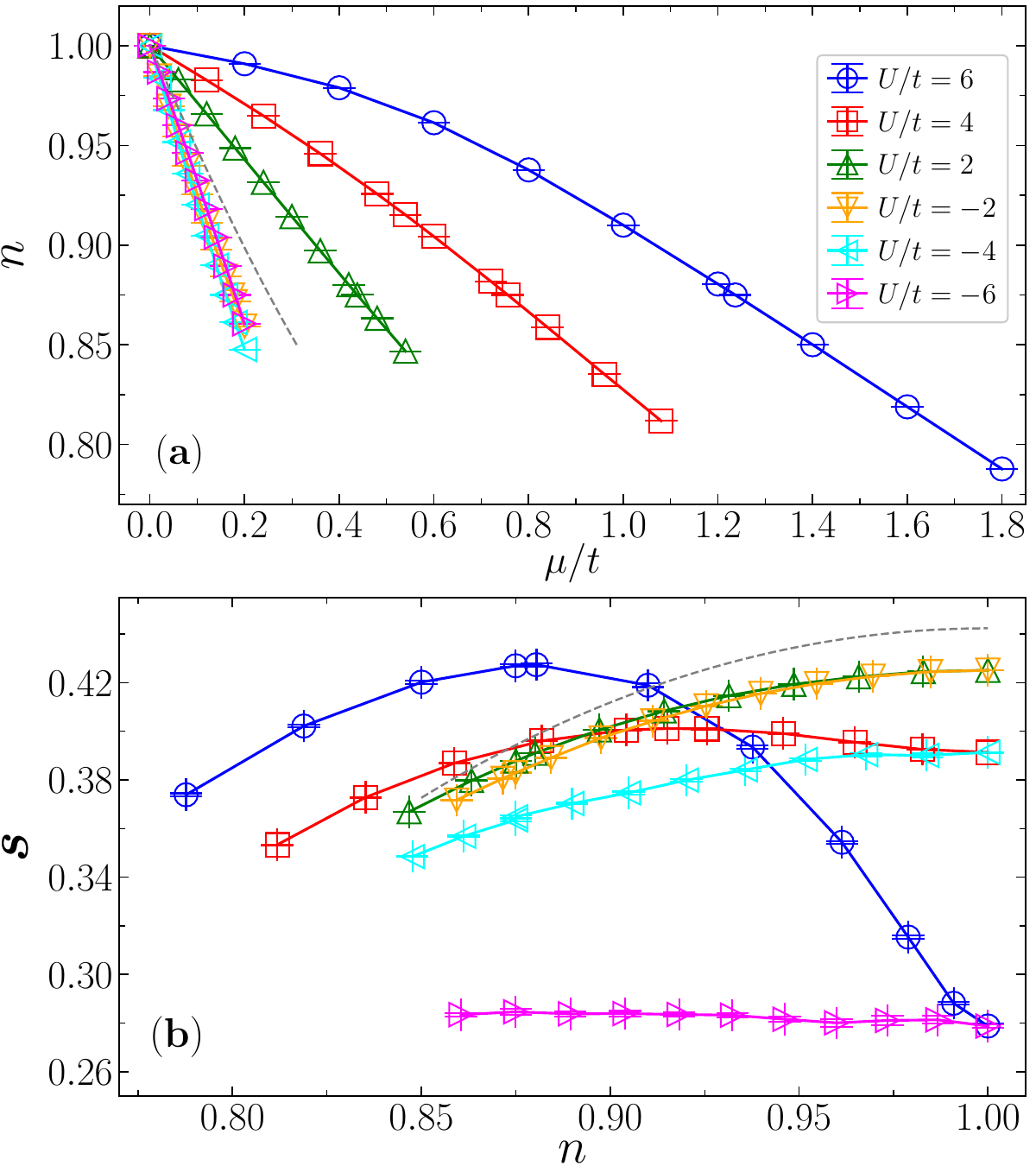}
\caption{Test results for the varying-$\mu$ path with fixed $U/t$ and $T/t=0.3$. (a) plots the fermion density $n$ as a function of $\mu/t$, and (b) shows the entropy density $\eyps$ vs $n$, computed via Eq.~(\ref{eq:changem}). The calculations are performed on an $L=4$ system for the 3D Hubbard model with $U/t=+6$, $+4$, $+2$, $-2$, $-4$, $-6$. The dashed lines in (a) and (b) denote the noninteracting results.}
\label{fig:Fig06EntropyPath3}
\end{figure}

\begin{figure}[t]
\centering
\includegraphics[width=0.953\columnwidth]{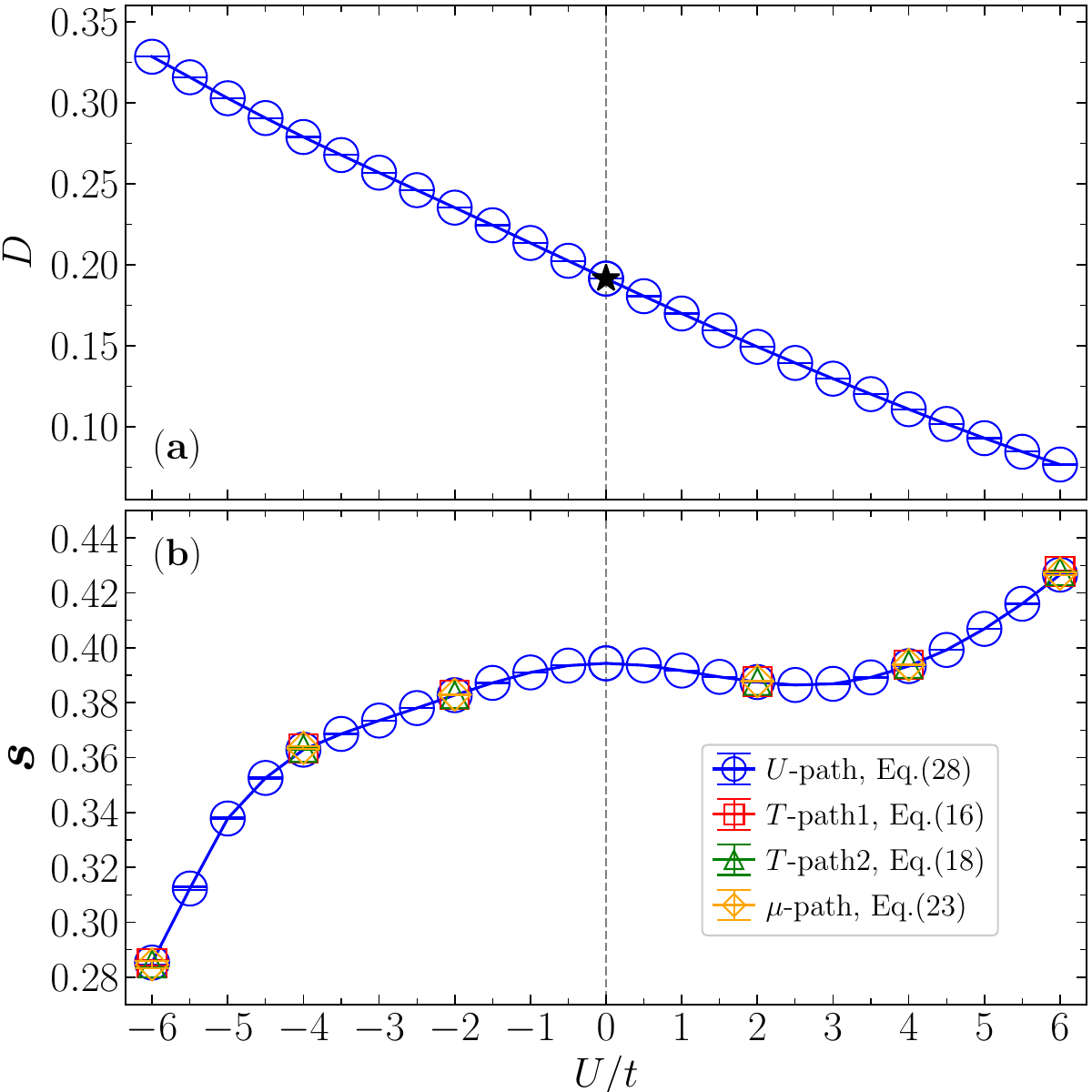}
\caption{Test results for the varying-$U$ path with fixed $T/t=0.3$ and $n=0.875$, and cross-check against the other three calculation schemes. (a) plots the double occupancy $D$ as a function of $U/t$ in the range $-6\le U/t\le +6$. The black star at $U/t=0$ marks the noninteracting result of $D=(7/16)^2$. (b) shows the entropy density $\eyps$ computed via Eq.~(\ref{eq:changeUfixTn1}) (blue circles, $U$-path). In (b), the $\eyps$ results at $(T/t=0.3$, $n=0.875)$ and $U/t=+6$, $+4$, $+2$, $-2$, $-4$, $-6$ from Fig.~\ref{fig:Fig03EntropyPath1} (red squares, $T$-path1), Fig.~\ref{fig:Fig04EntropyPath2} (green triangles, $T$-path2), and Fig.~\ref{fig:Fig06EntropyPath3} (orange diamonds, $\mu$-path) are also included for the check across different calculation schemes for the entropy. The calculations are performed on an $L=4$ system for the 3D Hubbard model.}
\label{fig:Fig07EntropyPath4}
\end{figure}

For the varying-$\mu$ path with fixed $T/t$ and $U/t$, we plot the test results in Fig.~\ref{fig:Fig06EntropyPath3}, including fermion density $n$ and entropy density $\eyps$ computed via Eq.~(\ref{eq:changem}). The calculations are performed at $T/t=0.3$ for six values of $U/t$. We observe that, for both $U>0$ and $U<0$, the fermion density $n$ decreases very smoothly with $\mu$ and it can always achieve high precision. Note $\mu=0$ represents half-filling case with $n=1$ in the model~(\ref{eq:Hamiltonian}). The variation of $n$ with $\mu$ involves the charge compressibility $\kappa=-\partial n/\partial\mu$, which is inversely correlated with the charge gap of the model at $T=0$. This leads to the relatively slower variation in $n(\mu)$ curve with increasing $U/t$ for $U>0$. As a comparison, for $U/t<0$, $n$ is rapidly suppressed by $\mu$, indicating large $\kappa$, which should be related to the superconducting ground state of the attractive Hubbard model. In Fig.~\ref{fig:Fig06EntropyPath3}(b), we directly plot $\eyps$ as a function of $n$, using the results in Fig.~\ref{fig:Fig06EntropyPath3}(a). We find that, for $U/t=-2$ and $-4$, $\eyps$ slightly decreases with the doping ($\delta=1-n$), whereas it remains nearly unchanged at $U/t=-6$. These behaviors call for further investigation with large-scale simulations. For $U/t>0$, $\eyps(n)$ at $U/t=2$ gradually traces the noninteracting curve, and interestingly exhibits a nonmonotonic dependence on $n$ at $U/t=4$ and $6$. The latter is related to the essential physics of the doped Hubbard model and will be systematically analyzed in Sec.~\ref{Sec:PhysicsIn{2D}}. Moreover, these $\eyps(n)$ results roughly suggest $(\partial\eyps/\partial n)_{U,T}=0$ at $n=1$, consistent with the Maxwell relation in Eq.~(\ref{eq:Maxwell000}), since $\mu=0$ is $T$-independent at half filling.

\begin{figure}[t]
\centering
\includegraphics[width=1.00\columnwidth]{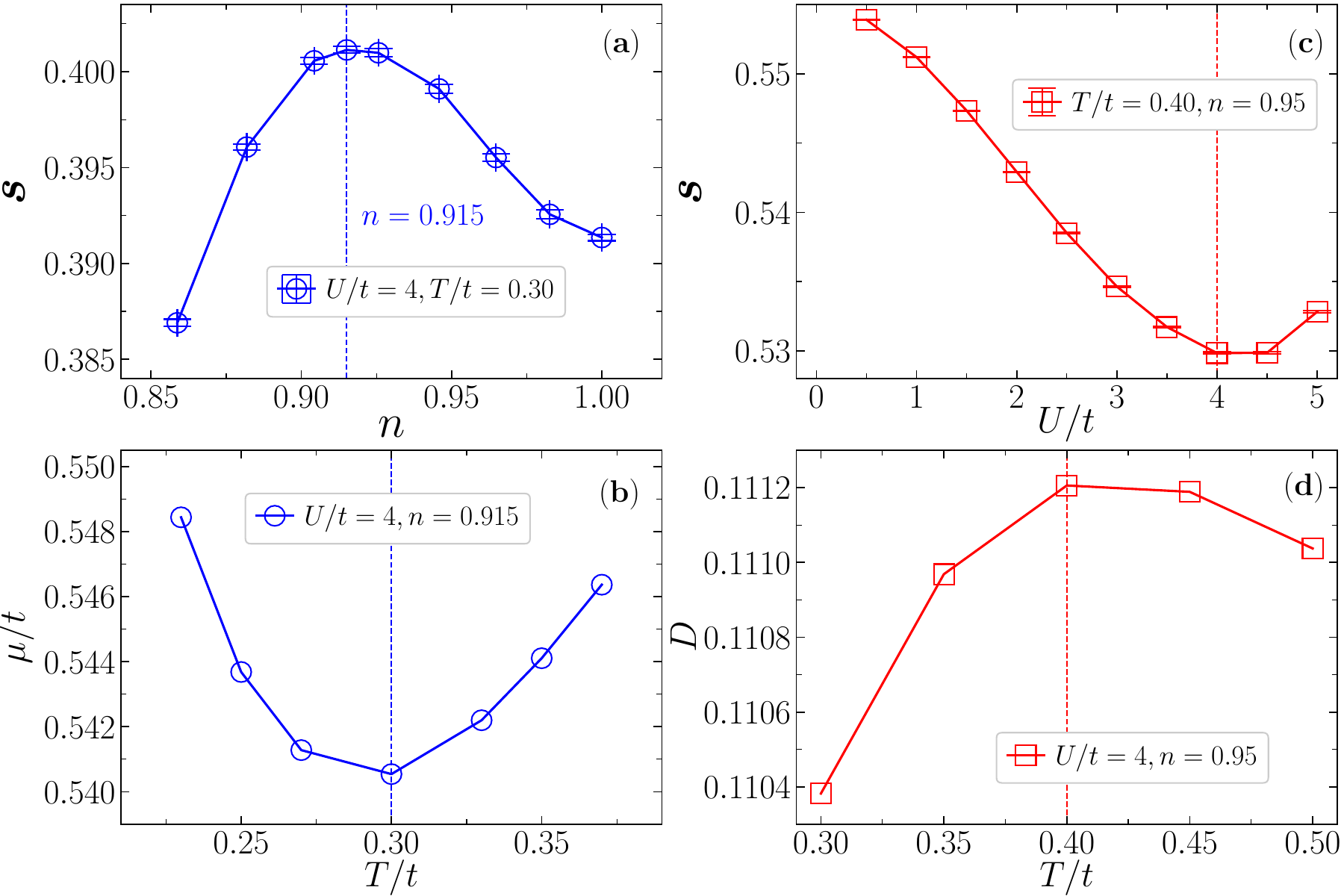}
\caption{Illustration of the Maxwell relations in Eqs.~(\ref{eq:Maxwell000}) and~(\ref{eq:Maxwell002}). At fixed $U/t=4$, (a) shows the entropy density $\eyps$ at $T/t=0.3$ vs fermion density $n$ tuned by the chemical potential $\mu$, while (b) plots the determined $\mu/t$ vs $T/t$ with fixed $n=0.915$. At fixed $n=0.95$, (c) plots $\eyps$ at $T/t=0.4$ vs $U/t$, while (d) presents the double occupancy $D$ vs $T/t$ with fixed $U/t=4$. The vertical dashed lines in each panel mark the extremum position of the curve. The calculations are performed on an $L=4$ system for the 3D Hubbard model.}
\label{fig:Fig08Maxwell}
\end{figure}

For the varying-$U$ path with $T/t=0.3$ and $n=0.875$, we present the test results in Fig.~\ref{fig:Fig07EntropyPath4}, including double occupancy $D$ and entropy density $\eyps$ computed via Eq.~(\ref{eq:changeUfixTn1}). As illustrated in Fig.~\ref{fig:Fig07EntropyPath4}(a), $D(U)$ varies smoothly across both the attractive and repulsive regimes and is highly precise, facilitating the numerical evaluation of the integral in Eq.~(\ref{eq:changeUfixTn1}). In AFQMC simulations, we find that the high precision of $D$ makes its Trotter error more noticeable than in other observables (see Appendix~\ref{Sec:A1douocc}). Accordingly, we set $\Delta\tau t=0.015$ in these varying-$U$ calculations. The resulting $\eyps$ in Fig.~\ref{fig:Fig07EntropyPath4}(b) decreases with $|U|/t$ at $U<0$ side, whereas it exhibits a nonmonotonic dependence on $U$ for $U>0$. These nontrivial behaviors need to be verified in future large-scale simulations. The $\eyps$ results also conform with $(\partial\eyps/\partial U)_{n,T}=0$ at $U=0$ point, which is evident from the Maxwell relation in Eq.~(\ref{eq:Maxwell002}). Furthermore, to enable a direct comparison across different evaluation schemes for the entropy, we plot the $\eyps$ results at $(T/t=0.3$, $n=0.875)$ and $U/t=+6$, $+4$, $+2$, $-2$, $-4$, $-6$ obtained from the other three calculation schemes in Fig.~\ref{fig:Fig07EntropyPath4}(b). They include the fixed-$T$ paths in Figs.~\ref{fig:Fig03EntropyPath1} and~\ref{fig:Fig04EntropyPath2} and the fixed-$\mu$ path in Fig.~\ref{fig:Fig06EntropyPath3}. All four schemes yield perfectly consistent results for $\eyps$ in the doped Hubbard model, thereby validating our computational framework. This agreement further suggests that, in practical calculations, one may apply the most convenient scheme to evaluate the entropy at a given point $(T/t, U/t, n)$ in the parameter space of the Hubbard model.

\begin{figure*}
\centering
\includegraphics[width=1.95\columnwidth]{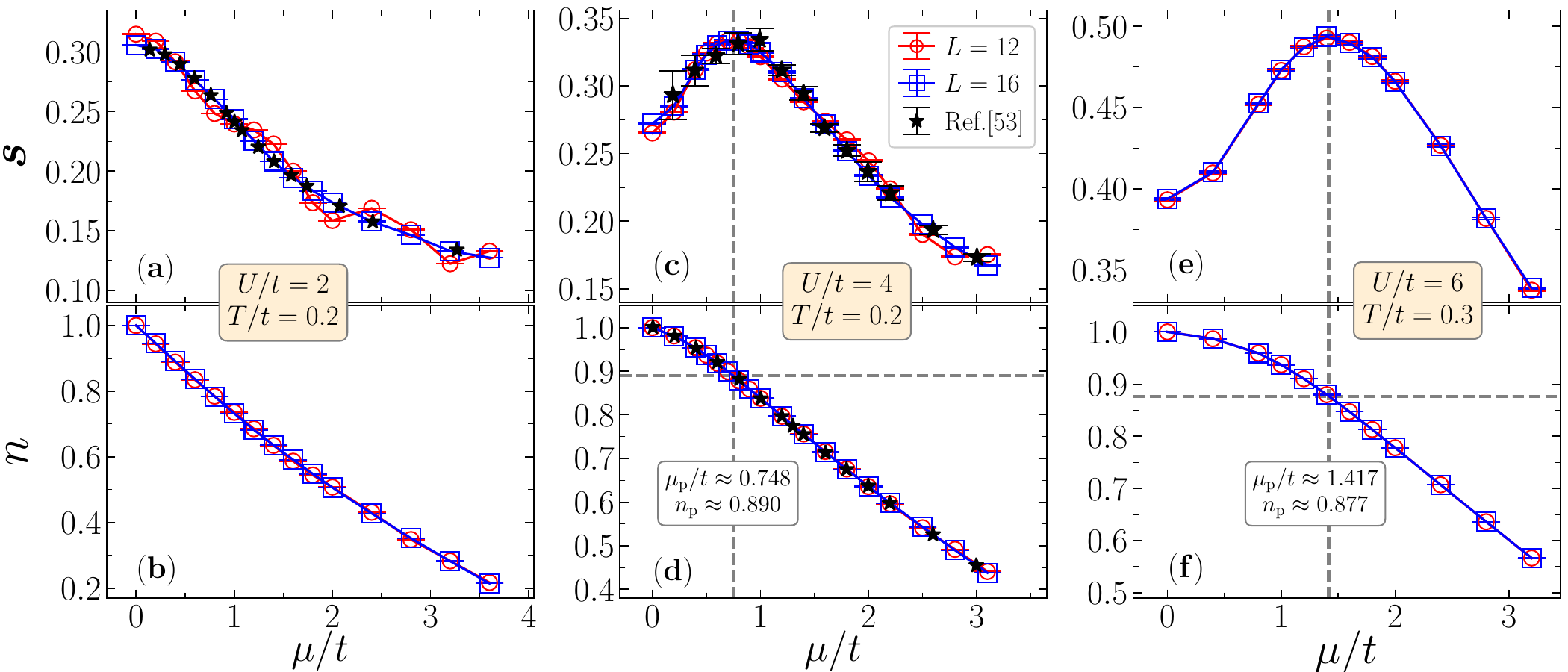}
\caption{Entropy density $\eyps$ (top panels) and fermion density $n$ (bottom panels) as functions of chemical potential $\mu/t$ for the {2D} Hubbard model from $L=12$ (red circles) and $L=16$ (blue squares) systems. (a) and (b) show results for $U/t=2$, $T/t=0.2$; (c) and (d) for $U/t=4$, $T/t=0.2$; and (e) and (f) for $U/t=6$, $T/t=0.3$. Black stars denote previous DiagMC results from Ref.~\cite{Lenihan2021}. In (c) and (e), gray vertical dashed lines mark the peak positions of $\eyps$, whose corresponding $\mu$ and $n$ are indicated in (d) and (f) by the dashed lines, with their values labeled.}
\label{fig:Fig09EntropyPhys}
\end{figure*}

We then turn to the numerical check of the Maxwell relations in Eqs.~(\ref{eq:Maxwell000}) and~(\ref{eq:Maxwell002}) for the doped Hubbard model. The test results are presented in Fig.~\ref{fig:Fig08Maxwell}. According to Eq.~(\ref{eq:Maxwell000}), with fixed $U$, the extremum position of the $\eyps(n)$ curve [as $(\partial\eyps/\partial n)_{U,T}=0$] coincides with that of the $\mu(T)$ curve [as $(\partial\mu/\partial T)_{U,n}=0$]. This correspondence is apparently satisfied in Figs.~\ref{fig:Fig08Maxwell}(a) and~\ref{fig:Fig08Maxwell}(b), where the position is around $(T/t=0.30,n=0.915)$. This extremum position depends on $T/t$ and $U/t$. Similarly, Eq.~(\ref{eq:Maxwell002}) means that, at fixed $n$, the $\eyps(U)$ curve and $D(T)$ curve have the same extremum position on the $T$-$U$ plane. This is also confirmed in Figs.~\ref{fig:Fig08Maxwell}(c) and~\ref{fig:Fig08Maxwell}(d), where the extremum position is around $(U/t=4,T/t=0.30)$ for the specific parameters applied. A complete numerical verification of these Maxwell relations requires the full landscapes of $\eyps$ and $\mu$ on the $T$-$n$ plane for Eq.~(\ref{eq:Maxwell000}), and those of $\eyps$ and $D$ on the $T$-$U$ plane for Eq.~(\ref{eq:Maxwell002}).

\subsection{Physically relevant results in the {2D} doped Hubbard model}
\label{Sec:PhysicsIn{2D}}

The test results in the previous subsection demonstrate the effectiveness and reliability of our computational framework for evaluating the thermal entropy. We now focus on the physically relevant parameter regime of the {2D} doped Hubbard model and elucidate its essential physics based on AFQMC calculations of the entropy. We apply $\Delta\tau t=0.04$ in the simulations. 

In Fig.~\ref{fig:Fig09EntropyPhys}, we present the results from varying-$\mu$ calculations with fixed $U/t$ and $T/t$, showing the thermal entropy density $\eyps$ and fermion density $n$ as functions of $\mu/t$. The simulations are carried out for the parameter sets $(U/t=2$, $T/t=0.2)$, $(U/t=4$, $T/t=0.2)$, and $(U/t=6$, $T/t=0.3)$, for which the fermion sign problem remains moderate (see Appendix~\ref{Sec:A1douocc}). Comparing the results for $L = 12$ and $16$, it is clear that the $n(\mu)$ data have already converged to the thermodynamic limit for all three parameter sets. Here $\eyps$ is computed via Eq.~(\ref{eq:changem}), where the integral of $n(\mu)$ can be evaluated numerically with high efficiency and accuracy due to the rather smooth behavior of $n(\mu)$. The observable oscillation in the $\eyps$ results for $U/t=2$ and $4$ at $L=12$ is attributed to nonsmoothness of $e_{\rm tot}(\mu)$, due to the single-particle finite-size effect~\cite{Song2025B}. This issue almost disappears in $L=16$ results. Moreover, we benchmark our results against those from the previous DiagMC study~\cite{Lenihan2021}, where the entropy was evaluated from the directly computed grand potential. Excellent agreement is found between the two calculations for $U/t=2$ and $4$, further confirming the validity of our entropy evaluation schemes. However, the DiagMC calculations in Ref.~\cite{Lenihan2021} show large error bars in $\eyps$ even for $U/t=4$ and do not access the $U/t=6$ case due to convergence issues in the diagrammatic series. In contrast, our AFQMC simulations provide substantially more precise data for $U/t=4$ and yield fully converged results for $U/t=6$ at $T/t=0.3$.

The most characteristic feature of $\eyps$ versus $\mu/t$ is the peak structure, clearly visible for $U/t=4$ and $6$ in Figs.~\ref{fig:Fig09EntropyPhys}(c) and~\ref{fig:Fig09EntropyPhys}(e). This peak in $\eyps$ also exists in the 3D doped Hubbard model even with $L=4$, as shown in Fig.~\ref{fig:Fig06EntropyPath3}(b). Such behavior of $\eyps$ versus $\mu/t$ or $n$ (or doping $\delta=1-n$) was first reported via dynamical mean-field theory (DMFT) calculations for the {2D} doped Hubbard model in Ref.~\cite{Khatami2009}. The study found that, with decreasing $T/t$, the corresponding filling $n_{\rm p}$ at the peak position $\mu_{\rm p}/t$ evolves toward the doping-induced quantum critical point (QCP) at $T=0$~\cite{Macridin2009,Khatami2009,Khatami2011}. A subsequent DMFT study~\cite{Sordi2011} further confirmed that $n_{\rm p}$ exhibits only weak temperature dependence. Later, this peak behavior in $\eyps(\mu)$ was also observed in optical lattice experiments~\cite{Cocchi2017}. In recent years, the ground-state properties of the {2D} system have been extensively studied using cutting-edge quantum many-body approaches~\cite{Qin2022,Mingpu2016,Boxiao2017,Qin2020,Haoxu2022}. These works suggest that the doping-induced QCP is likely associated with a quantum phase transition between the stripe-ordered phase and disordered metallic phase in the underdoped and overdoped regimes, respectively, of the {2D} doped Hubbard model (without next-nearest-neighbor hopping $t^{\prime}$). A similar doping-induced QCP has also been identified in the magnetic phase diagram of the 3D doped Hubbard model~\cite{Rampon2025,Katanin2017}.

Understanding the connection between the peak position of $\eyps(\mu)$ [or equivalently $\eyps(n)$] and the QCP in the doped Hubbard model relies on the Maxwell relations in Eqs.~(\ref{eq:Maxwell000}) and~(\ref{eq:Maxwellb}). The peak in $\eyps(\mu)$ at $\mu_{\rm p}/t$ implies a corresponding peak in $\eyps(n)$ at $n_{\rm p}=n(\mu_{\rm p})$, featuring $(\partial\eyps/\partial n)_{U,T}=0$ at $n=n_{\rm p}$ and $(\partial\eyps/\partial n)_{U,T}<0$ at $n>n_{\rm p}$ [and oppositely $(\partial\eyps/\partial n)_{U,T}>0$ at $n<n_{\rm p}$]. According to Eq.~(\ref{eq:Maxwell000}), these relations translate into
\begin{subequations}\begin{align}
&\Big(\frac{\partial \mu}{\partial T}\Big)_{U,n} = 0\ \ \ {\rm at}\ \ n=n_{\rm p}, \label{eq:PmuPTa} \\
&\Big(\frac{\partial \mu}{\partial T}\Big)_{U,n} < 0\ \ \ {\rm for}\ \ n>n_{\rm p}, \label{eq:PmuPTb}\\
&\Big(\frac{\partial \mu}{\partial T}\Big)_{U,n} > 0\ \ \ {\rm for}\ \ n<n_{\rm p}. \label{eq:PmuPTc}
\end{align}\end{subequations}
The first equality indicates a stationary chemical potential at $n=n_{\rm p}$, which can be interpreted as the signature of local particle-hole symmetry~\cite{Macridin2009,Khatami2009} in analogy to the half-filled case. Such near particle-hole symmetry has indeed been observed in the cuprates around optimal doping~\cite{Galanakis2010}. In addition, Eqs.~(\ref{eq:PmuPTb}) and~(\ref{eq:PmuPTc}) reveal totally different low-$T$ behaviors of $\mu(T)$ (at fixed $n$) in the regimes $n>n_{\rm p}$ and $n<n_{\rm p}$, namely $\mu(T)$ increases or decreases, respectively, with decreasing $T/t$. From a theoretical perspective~\cite{Khatami2011}, the former [Eq.~(\ref{eq:PmuPTb})] conforms with the pseudogap state, which has been shown to evolve continuously into the $T=0$ stripe-ordered phase as $T/t$ decreases in the {2D} doped Hubbard model~\cite{Fedor2024}. And the latter [Eq.~(\ref{eq:PmuPTc})] aligns with the Fermi-liquid state. Taken together, these features elucidate that $n_{\rm p}$, the peak filling of $\eyps(n)$, evolves into the critical filling at the doping-induced QCP as $T/t$ approaches zero. The same conclusion can also be reached via Eq.~(\ref{eq:Maxwellb}), which instead focuses on the $T$ dependence of the fermion density $n$ at fixed $\mu$. The condition $(\partial\eyps/\partial\mu)_{U,T}=(\partial n/\partial T)_{U,\mu}=0$ at $\mu=\mu_{\rm p}$, corresponding to a stationary $n$ with respect to $T$, likewise signals a local particle-hole symmetry. It separates two regimes $\mu<\mu_{\rm p}$ (with $n>n_{\rm p}$) and $\mu>\mu_{\rm p}$ (with $n<n_{\rm p}$), which feature $(\partial n/\partial T)_{U,\mu}<0$ and $(\partial n/\partial T)_{U,\mu}>0$, respectively, according to Eq.~(\ref{eq:Maxwellb}). These opposite $T$ dependences of $n(T)$ at fixed $\mu$ distinguish two distinct states, namely pseudogap and Fermi liquid, in the low-$T$ regime of the doped Hubbard model.

As included in Fig.~\ref{fig:Fig09EntropyPhys}, we have determined the peak position of $\eyps(\mu)$ or $\eyps(n)$ as $(\mu_{\rm p}/t\simeq0.748$, $n_{\rm p}\simeq0.890)$ for $(U/t=4$, $T/t=0.2)$, and $(\mu_{\rm p}/t\simeq1.417$, $n_{\rm p}\simeq0.877)$ for $(U/t=6$, $T/t=0.3)$. At $U/t=6$, our result of $n_{\rm p}$ is slightly larger than the DMFT result as $n_{\rm p}\simeq 0.85$, reported in Ref.~\cite{Macridin2009,Khatami2009}. The difference may be attributed to the residual temperature dependence of $n_{\rm p}$ or the approximation introduced in DMFT method. For $U/t=2$, as shown in Fig.~\ref{fig:Fig09EntropyPhys}(a), the entropy $\eyps(\mu)$ decreases monotonically with $\mu$, exhibiting no peak. This behavior may arise because the QCP for $U/t = 2$ lies very close to half filling, and the temperature $T/t = 0.2$ used here is not low enough to reveal the features discussed above. Besides the peak behavior, $\eyps(\mu)$ also features an inflection point, corresponding to the point of most rapid increase with $\mu$, in the range $0<\mu<\mu_{\rm p}$. This is reflected by the equality $(\partial\eyps/\partial\mu)_{U,T}=0$ at both $\mu=\mu_{\rm p}$ and $\mu=0$, where the former is evident and the latter arises from the Maxwell relation in Eq.~(\ref{eq:Maxwellb}). The inflection point of $\eyps(\mu)$ can be determined by $\partial^2\eyps(\mu)/\partial\mu^2=0$, and it actually separates two regimes characterized by $\partial\kappa/\partial T<0$ and $\partial\kappa/\partial T>0$, respectively, according to Eq.~(\ref{eq:ParSParMu2nd}). In Ref.~\cite{Lenihan2021}, this inflection point is used to identify the crossover from a non-Fermi-liquid state near half filling to the metallic state with larger doping. For the numerical results, we obtain the inflection point of $\eyps(\mu)$ at $(\mu/t\simeq0.32$, $n\simeq0.959)$ for $(U/t=4$, $T/t=0.2)$ and at $(\mu/t\simeq0.75$, $n\simeq0.964)$ for $(U/t=6$, $T/t=0.3)$.

From the above analysis, the fermion filling (or doping) dependence of the thermal entropy can serve as an effective thermodynamic probe of the doping-induced QCP in the doped Hubbard model, while also providing valuable insights into finite-temperature crossover phenomena. Our calculation scheme, based on the varying-$\mu$ path with fixed $U$ and $T$ via Eq.~(\ref{eq:changem}), offers a highly efficient approach for mapping out the full landscape of thermal entropy in the temperature-filling plane. This may help elucidate the phase diagram of the doped Hubbard model, a long-standing problem in theoretical condensed matter physics.

\section{Summary and discussion}
\label{Sec:Summary}

In summary, we have established a unified and highly efficient framework for computing the thermal entropy in the doped Fermi-Hubbard model. Building on the fundamental quantity of the grand potential, we have achieved four complementary evaluation schemes obtained from two independent theoretical derivations. In these schemes, the entropy is expressed as simple path integrals over temperature, chemical potential, and interaction strength within the parameter space of the model. We have also derived useful Maxwell relations that link the thermal entropy to other thermodynamic quantities. As an early-stage application, we have employed numerically unbiased AFQMC method to systematically test the entropy calculations in the 3D Hubbard model and to perform cross-checks between different evaluation schemes. Furthermore, in the physically relevant parameter regime of the {2D} doped Hubbard model, our numerical results show nice agreement with previous DiagMC calculations, while also revealing a deep connection between characteristic features of the entropy landscape and the essential physics of the model. 

Our computational framework for the thermal entropy can have broader applications. First, it can be readily applied across a wide range of quantum many-body numerical methods. This is because the entropy admits simple expressions [see Eqs.~(\ref{eq:fixednT}),~(\ref{eq:fixedmT}),~(\ref{eq:changem}), and~(\ref{eq:changeUfixTn1})], and the integrands in these evaluation schemes only involve the total energy, fermion density, and double occupancy, all of which are straightforward to compute in most numerical approaches. Second, the framework is not restricted to the grand canonical ensemble simulations for the Hubbard model. Specifically, the varying-$T$ scheme in Sec.~\ref{Sec:FixUn} and the varying-$U$ scheme in Sec.~\ref{Sec:FixTn} can be applied to numerical methods working in the canonical ensemble, such as the finite-temperature canonical-ensemble AFQMC method~\cite{Gilbreth2021,Wan2021,Ding2025,Xu2026} and the minimally entangled typical thermal states method~\cite{White2009,Wietek2021,Wietek2022}. Third, the framework can be further extended to fit into generalized Hubbard models. For example, in the {2D} Hubbard model with next-nearest-neighbor hopping $t^{\prime}$~\cite{Jiang2020,Haoxu2024,Zhang2025} and the 3D Hubbard model with hopping anisotropy in $z$ direction (denoted as $t_z$)~\cite{Troyer2014,Ibarra2020}, one can follow the formalism in Sec.~\ref{Sec:Ensemble}, add $t^{\prime}$ or $t_z$ as a new natural variable to the grand potential, and reach a similar scheme to calculate the entropy as a function of the variable. Or equivalently, one can set $\alpha=t^{\prime}$ (or $\alpha=t_z$) in Eq.~(\ref{eq:HFfiniteT}) to get the corresponding Hellmann-Feynman relation and then achieve the expression of thermal entropy. Our computational framework, together with these promising generalizations, can help broaden the scope and deepen the study of the doped Hubbard models from a thermodynamic perspective centered on thermal entropy.

\begin{acknowledgments}
This work was supported by the National Natural Science Foundation of China (Grant Nos. 12247103, 12204377, and 12275263), the Quantum Science and Technology-National Science and Technology Major Project (Grant No. 2021ZD0301900), the Natural Science Foundation of Fujian province of China (Grant No. 2023J02032), and the Youth Innovation Team of Shaanxi Universities.
\end{acknowledgments}

\appendix
\section{The Trotter error in double occupancy, and the sign average for Fig.~\ref{fig:Fig09EntropyPhys}}
\label{Sec:A1douocc}

In this appendix, we show additional AFQMC results for the Trotter errors in double occupancy and the corresponding thermal entropy, as well as for the fermion sign average for the results shown in Fig.~\ref{fig:Fig09EntropyPhys} in the main text. 

\begin{figure}
\centering
\includegraphics[width=0.98\columnwidth]{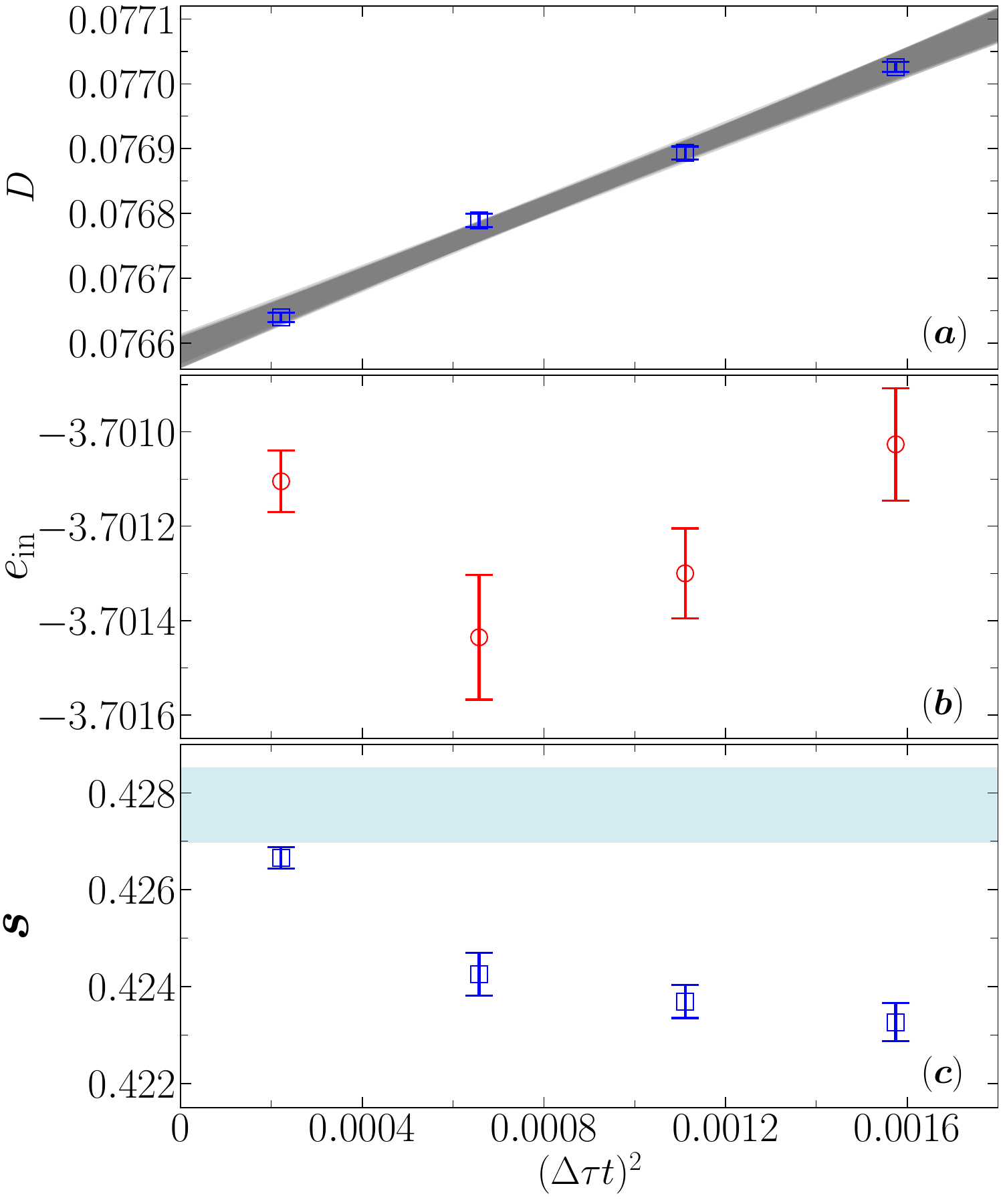}
\caption{Trotter error tests in double occupancy $D$ and thermal entropy density $\eyps$ computed via Eq.~(\ref{eq:changeUfixTn1}). (a) and (b) plot $D$ and the energy density $e_{\rm in}/t$ as a function of $(\Delta\tau t)^2$ at $T/t=0.3$ and $U/t=6$ with fixed filling $n=0.875$, where $\Delta\tau$ is the Trotter time step defined as $\Delta\tau=\beta/M$. In (a), the gray band represents the linear fitting versus $(\Delta\tau t)^2$ for the $D$ results. (c) shows the results of $\eyps$, while the light blue band denotes the $\eyps$ result evaluated via Eq.~(\ref{eq:fixednT}) along the varying-$T$ path with fixed $U$ and $n$. The calculations are performed on an $L=4$ system for the 3D Hubbard model at $T/t=0.3$ and $U/t=6$ with fixed filling $n=0.875$.}
\label{fig:A1douocc}
\end{figure}

\begin{figure*}
\centering
\includegraphics[width=1.95\columnwidth]{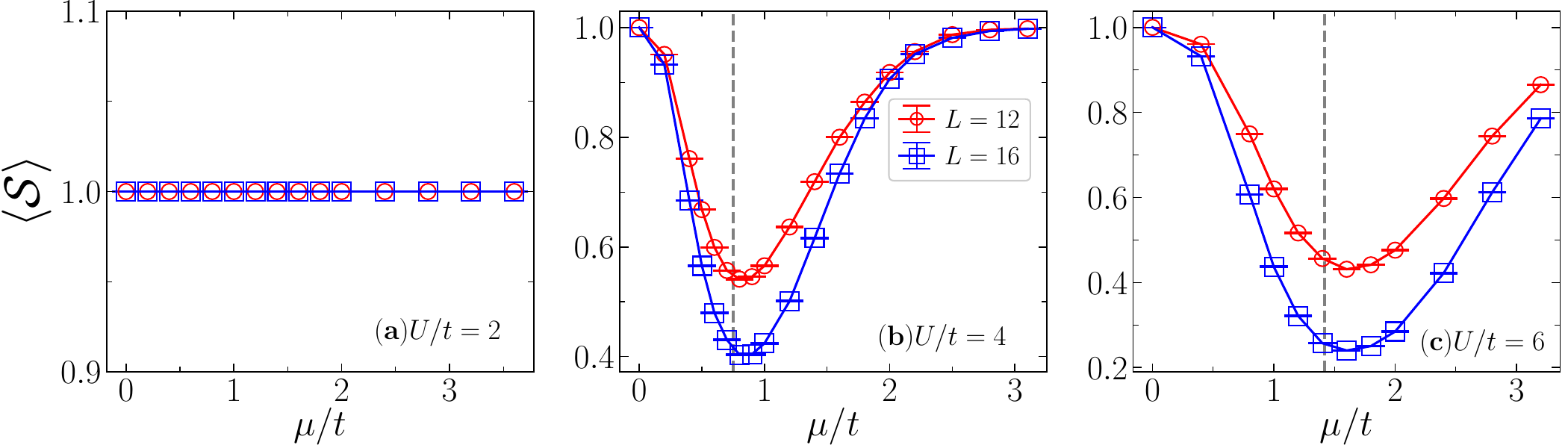}
\caption{The average sign $\langle\mathcal{S}\rangle$ of the corresponding AFQMC simulations in Fig.~\ref{fig:Fig09EntropyPhys} in the main text. The results are presented for three parameter sets as (a) $U/t=2$, $T/t=0.2$, (b) $U/t=4$, $T/t=0.2$, and (c) $U/t=6$, $T/t=0.3$. In (b) and (c), gray vertical dashed lines mark the peak positions of $\eyps(\mu)$, as $\mu_{\rm p}/t\simeq0.748$ for $U/t=4$ case and $\mu_{\rm p}/t\simeq1.417$ for $U/t=6$ case. }
\label{fig:A2SignAverage}
\end{figure*}

In Sec.~\ref{Sec:FixTn}, we show that the entropy can be evaluated via the varying-$U$ path with fixed $T$ and $n$ [see Eq.~(\ref{eq:changeUfixTn1})], in which the integral of double occupancy $D$ over $U$ is involved. During the testing calculations in Sec.~\ref{Sec:Testin3D}, we find that the AFQMC results of $D$ exhibit substantial Trotter error. This is actually due to the fact that the precision of $D$ is significantly higher than many other observables, thus rendering its Trotter error more noticeable within the same $\Delta\tau$. Figure~\ref{fig:A1douocc} presents the test results of Trotter error in both $D$ and thermal entropy density $\eyps$ for the 3D Hubbard model at $T/t=0.3$, $U/t=6$, and $n=0.875$. The four data points correspond to $\Delta\tau t = 0.015, 0.026, 0.034$ and $0.040$ in the AFQMC simulations. Note the extremely small scales in all three panels. The relative errors for the $D$ results in Fig.~\ref{fig:A1douocc}(a) are smaller than or close to $0.01\%$. We find that $D$ versus $(\Delta\tau t)^2$ fits well to a linear dependence, as expected for the symmetric Trotter decomposition applied in our AFQMC calculations. As illustrated, $D(\Delta\tau t=0.015)$ is roughly consistent with the extrapolated result of $D$ at $\Delta\tau t=0$ considering the statistical uncertainties. For the energy density $e_{\rm in}$ shown in Fig.~\ref{fig:A1douocc}(b), the results are consistent within twice the error bars, indicating that the Trotter error is nearly eliminated. The slight variations may be induced by small differences in the fermion density during the fixed-$n$ procedure. To assess the Trotter error in $\eyps$, we repeat the AFQMC simulations for each $\Delta\tau t$ over the range $0 \le U/t \le 6$, and compute the corresponding $\eyps$ via Eq.~(\ref{eq:changeUfixTn1}). The resulting $\eyps$, shown in Fig.~\ref{fig:A1douocc}(c), also exhibits noticeable Trotter errors and converges to the result obtained from the varying-$T$ calculations at fixed $U$ and $n$ [see Eq.~(\ref{eq:fixednT})], which we have verified to be essentially free of Trotter error. The largest Trotter error in $\eyps$ at $\Delta\tau t=0.04$ is about $1\%$ (relative). Based on the above test results, we find that, at $\Delta\tau t=0.015$, the Trotter errors in both $D$ and $\eyps$ are negligible. Hence we apply $\Delta\tau t=0.015$ for the varying-$U$ calculations in Sec.~\ref{Sec:Testin3D}. 

In Fig.~\ref{fig:A2SignAverage}, we present the results of the average sign $\langle\mathcal{S}\rangle$ of the corresponding AFQMC simulations in Fig.~\ref{fig:Fig09EntropyPhys} in the main text. For the parameter set $(U/t=2$, $T/t=0.2)$, $\langle\mathcal{S}\rangle$ remains close to unity for both $L=12$ and $16$, indicating a negligible sign problem. For the cases of $U/t=4$ and $6$, a significant sign problem is observed, with the minimum $\langle\mathcal{S}\rangle$ reaching $\sim 0.4$ for $U/t=4$ and $\sim 0.2$ for $U/t=6$. Interestingly, for both cases, $\langle\mathcal{S}\rangle$ reaches its minimum around the peak position ($\mu_{\rm p}$) of $\eyps(\mu)$, which, according to our discussion in Sec.~\ref{Sec:PhysicsIn{2D}}, is associated with the doping-induced quantum critical point (QCP) at $T=0$. Hence, the most severe sign problem in the doped Hubbard model is likely correlated with the QCP, consistent with previous observations of a particularly severe sign problem near optimal doping~\cite{Iglovikov2015} and with the general connection between the average sign and quantum critical behavior reported in Ref.~\cite{Mondaini2022}.

\section{Derivations for the entropy at certain limits and the proof of Eqs.~(\ref{eq:Maxwell002}),~(\ref{eq:AtomicD}), and \texorpdfstring{$\eyps(+\mu)=\eyps(-\mu)$}{}}
\label{Sec:A2Entropy}

In this appendix, we focus on the detailed derivations for some formulas or expressions used in the main text. First, we derive the expressions of the thermal entropy in noninteracting systems as well as at $T=\infty$ with fixed fermion filling. Second, we present the proof for Eqs.~(\ref{eq:Maxwell002}) and~(\ref{eq:AtomicD}). Third, we prove $\eyps(+\mu)=\eyps(-\mu)$ for the model~(\ref{eq:Hamiltonian}) using Eq.~(\ref{eq:changem}).

For the noninteracting limit ($U=0$) of the model~(\ref{eq:Hamiltonian}), we can compute its partition function $Z={\rm Tr}(e^{-\beta\hat{H}})$, and accordingly obtain the grand potential density via $\gpps=-N_s^{-1}T\ln Z$. This yields the following result
\begin{equation}\begin{aligned}
\Omega = - N_s^{-1} T \sum_{\mathbf{k}, \sigma}{\rm ln}\big[ 1 + e^{-\beta(\varepsilon_{\mathbf{k}\sigma}+\mu)} \big],
\end{aligned}\end{equation}
where $\varepsilon_{\mathbf{k}\sigma}$ is the kinetic energy dispersion, with its specific expression in both {2D} and 3D given in Sec.~\ref{sec:ModelMethod}. Then the entropy density follows from the thermodynamic relation
\begin{equation}\begin{aligned}
\label{eq:U0entropy}
\eyps_0 
&= -\Big( \frac{\partial \Omega}{\partial T} \Big)_{U, \mu} \\
&= -N_s^{-1}\sum_{\mathbf{k},\sigma}\big[f_{\mathbf{k}\sigma} \ln f_{\mathbf{k}\sigma}+(1-f_{\mathbf{k}\sigma})\ln(1-f_{\mathbf{k}\sigma})\big],
\end{aligned}\end{equation}
where $f_{\mathbf{k}\sigma}=1/[e^{\beta(\varepsilon_{\mathbf{k}\sigma}+\mu)}+1]$ is the Fermi-Dirac distribution. The equality can be simplified to Eq.~(\ref{eq:U0Exact}) considering $\varepsilon_{\mathbf{k},\uparrow}=\varepsilon_{\mathbf{k},\downarrow}$ in the model~(\ref{eq:Hamiltonian}). 

For the system at $T=\infty$ with fixed $U$ and $n$, we can compute its entropy density via the result in Eq.~(\ref{eq:U0entropy}). At fixed $n$, the corresponding $\mu(T)$ can be obtained via solving the equation
\begin{equation}\begin{aligned}
n=\frac{1}{N_s}\sum_{\mathbf{k},\sigma} f_{\mathbf{k}\sigma}\!\left[\mu(T)\right].
\end{aligned}\end{equation}
Then substituting the resulting $\mu(T)$ into Eq.~(\ref{eq:U0entropy}) yields the exact noninteracting entropy. At $T=\infty$ (and $\beta=0$), any finite $U$ becomes irrelevant because it only enters into the calculations via $\beta U$ which is also zero. Therefore the $T\to\infty$ entropy at fixed $n$ is identical to that of the noninteracting system, provided that the correct $\mu$ is used to enforce the desired density. Starting from the Fermi-Dirac distribution,
\begin{equation}\begin{aligned}
f_{\mathbf{k}\sigma}(\mu)=\frac{1}{e^{\beta(\varepsilon_{\mathbf{k}\sigma}+\mu)}+1},
\end{aligned}\end{equation}
we note that, as $\beta\to 0$ the dispersion becomes negligible in the exponent, the occupation (function value) approaches a $\mathbf{k}$-independent result as
\begin{equation}\begin{aligned}
f_{\mathbf{k}\sigma}(\mu) \xrightarrow[\beta\to 0]{}\frac{1}{1+e^{\beta\mu}}.
\end{aligned}\end{equation}
Then the fixed-$n$ condition presents the constraint
\begin{equation}\begin{aligned}
n = \frac{1}{N_s}\sum_{\mathbf{k}\sigma} f_{\mathbf{k}\sigma}\xrightarrow[\beta\to 0]{} \frac{2}{1+e^{\beta \mu}},
\end{aligned}\end{equation}
which yields $f_{\mathbf{k}\sigma}=n/2$ at $T=\infty$ and also
\begin{equation}\begin{aligned}
\beta\mu \xrightarrow[\beta\to 0]{}\ln\!\left(\frac{2-n}{n}\right),\hspace{0.3cm}
\mu(T) \xrightarrow[T\to\infty]{}T \ln\!\left(\frac{2-n}{n}\right).\label{eq:mu_infty_fixedn}
\end{aligned}\end{equation}
This expression explicitly shows that $\mu(T)$ diverges as $T\to\infty$ for fixed fermion filling $n\ne1$. By substituting $f_{\mathbf{k}\sigma}=n/2$ into Eq.~(\ref{eq:U0entropy}), we obtain the $T=\infty$ entropy density at fixed $n$ as
\begin{equation}\begin{aligned}
\eyps_n(\infty)
&= -2\Big[\frac{n}{2}\ln\!\left(\frac{n}{2}\right) +\left(1-\frac{n}{2}\right)\ln\!\left(1-\frac{n}{2}\right)\Big] \\
&=\ln 4 - n\ln n - (2-n)\ln(2-n),
\end{aligned}\end{equation}
which is the result mentioned in Sec.~\ref{Sec:FixUn}.

We then turn to prove the Maxwell relation in Eq.~(\ref{eq:Maxwell002}) in the main text. First, with tuning $U$ at fixed $T$ and $n$, the chemical potential $\mu$ is also a function of $U$, and thus the entropy can be written as $\eyps(U,T,\mu(U))$. Hence, the total differential of $\eyps$ follows
\begin{equation}\begin{aligned}
\dd\eyps 
&= \Big(\frac{\partial\eyps}{\partial U}\Big)_{T,\mu}\dd U + \Big(\frac{\partial\eyps}{\partial\mu}\Big)_{T,U}\dd\mu \\
&= \Big(\frac{\partial\eyps}{\partial U}\Big)_{T,\mu}\dd U + \Big(\frac{\partial\eyps}{\partial\mu}\Big)_{T,U}\Big(\frac{\partial\mu}{\partial U}\Big)_{T,n}\dd U,
\end{aligned}\end{equation}
which leads to the equality
\begin{equation}\begin{aligned}
\Big(\frac{\partial\eyps}{\partial U}\Big)_{T,n}
= \Big(\frac{\partial\eyps}{\partial U}\Big)_{T,\mu} + \Big(\frac{\partial\eyps}{\partial\mu}\Big)_{T,U}\Big(\frac{\partial\mu}{\partial U}\Big)_{T,n}.
\end{aligned}\end{equation}
Using the equalities in Eqs.~(\ref{eq:Maxwella}),~(\ref{eq:Maxwellb}), and~(\ref{eq:Maxwell001}), the above equality can be reformulated as
\begin{equation}\begin{aligned}
\label{eq:PsPu}
\Big(\frac{\partial\eyps}{\partial U}\Big)_{T,n}
= -\Big(\frac{\partial h_{I}}{\partial T}\Big)_{U,\mu} + \Big(\frac{\partial n}{\partial T}\Big)_{U,\mu}\frac{\big(\frac{\partial n}{\partial U}\big)_{T,\mu}}
{\big(\frac{\partial n}{\partial\mu}\big)_{T,U}}.
\end{aligned}\end{equation}
Similarly, when tuning $T$ at fixed $U$ and $n$, the interaction energy density $h_{I}$ can be written as $h_{I}(U,T,\mu(T))$. Its total differential reads
\begin{equation}\begin{aligned}
\dd h_{I}
&= \Big(\frac{\partial h_{I}}{\partial T}\Big)_{U,\mu}\dd T + \Big(\frac{\partial h_{I}}{\partial\mu}\Big)_{T,U}\dd\mu \\
&= \Big(\frac{\partial h_{I}}{\partial T}\Big)_{U,\mu}\dd T + \Big(\frac{\partial h_{I}}{\partial\mu}\Big)_{T,U}\Big(\frac{\partial\mu}{\partial T}\Big)_{U,n}\dd T,
\end{aligned}\end{equation}
which leads to the equality
\begin{equation}\begin{aligned}
\Big(\frac{\partial h_{I}}{\partial T}\Big)_{U,n}
= \Big(\frac{\partial h_{I}}{\partial T}\Big)_{U,\mu} + \Big(\frac{\partial h_{I}}{\partial\mu}\Big)_{T,U}\Big(\frac{\partial\mu}{\partial T}\Big)_{U,n}.
\end{aligned}\end{equation}
Using the equalities in Eqs.~(\ref{eq:Maxwellc}) and~(\ref{eq:Maxwell000}), the above equality can be rewritten as
\begin{equation}\begin{aligned}
\label{eq:PhIPT}
\Big(\frac{\partial h_{I}}{\partial T}\Big)_{U,n}
= \Big(\frac{\partial h_{I}}{\partial T}\Big)_{U,\mu} - \Big(\frac{\partial n}{\partial U}\Big)_{T,\mu}\frac{\big(\frac{\partial n}{\partial T}\big)_{U,\mu}}{\big(\frac{\partial n}{\partial \mu}\big)_{U,T}}.
\end{aligned}\end{equation}
Then by comparing Eqs.~(\ref{eq:PsPu}) and~(\ref{eq:PhIPT}), the following equality is evident as
\begin{equation}\begin{aligned}
\Big(\frac{\partial\eyps}{\partial U}\Big)_{n,T} 
= -\Big(\frac{\partial h_{I}}{\partial T}\Big)_{n,U} 
= -\Big(\frac{\partial D}{\partial T}\Big)_{n,U},
\end{aligned}\end{equation}
in which the relation $h_{I}=D-n/2$ is used in the second equality. This is the Maxwell relation in Eq.~(\ref{eq:Maxwell002}).

The proof of Eq.~(\ref{eq:AtomicD}) involves the atomic limit under the condition $\beta U\ll 1$. Under this limit, the local Hamiltonian can be solved exactly, which allows us to derive the expansion of the double occupancy at fixed filling $n$. The atomic-limit Hamiltonian is
\begin{equation}\begin{aligned}
\hat h = U \hat n_{\uparrow}\hat n_{\downarrow} + \mu(\hat n_{\uparrow}+\hat n_{\downarrow}), \label{eq:single_site_H_atomic}
\end{aligned}\end{equation}
with partition function
\begin{equation}\begin{aligned}
Z = 1 + 2 e^{-\beta \mu} + e^{-\beta(U+2\mu)}.
\end{aligned}\end{equation}
The filling $n$ and the double occupancy $D$ are computed as
\begin{equation}\begin{aligned}
n = \frac{2e^{-\beta\mu}+2e^{-\beta(U+2\mu)}}{Z}, \quad D = \frac{e^{-\beta(U+2\mu)}}{Z}.
\label{eq:atomic_n_D_appendix}
\end{aligned}\end{equation}
Therefore we can reach 
\begin{equation}\begin{aligned}
\frac{D}{n} = \frac{1}{2\left[1+e^{\beta(U+\mu)}\right]}.
\label{eq:D_over_n_intermediate}
\end{aligned}\end{equation}
From the expression of $n$ in Eq.~\eqref{eq:atomic_n_D_appendix}, one can also find
\begin{equation}\begin{aligned}
(2-n)e^{-\beta U}e^{-2\beta\mu}+2(1-n)e^{-\beta\mu}-n=0.
\end{aligned}\end{equation}
Solving this quadratic equation for $e^{-\beta\mu}$ presents
\begin{equation}\begin{aligned}
e^{-\beta\mu}=\frac{n-1+\Delta}{e^{-\beta U}(2-n)},
\label{eq:emu_atomic_exact}
\end{aligned}\end{equation}
where $\Delta=\sqrt{(n-1)^2-e^{-\beta U}n(n-2)}$. By substituting Eq.~\eqref{eq:emu_atomic_exact} into Eq.~\eqref{eq:D_over_n_intermediate}, we obtain
\begin{equation}\begin{aligned}
\frac{D}{n} = \frac{1}{2}\Big[1+\frac{2-n}{n-1+\Delta}\Big]^{-1}.
\label{eq:D_over_n_exact}
\end{aligned}\end{equation}
Now we expand this expression at the limit $\beta U\to 0$.
Using the relation $e^{-\beta U}=1-\beta U+\mathcal{O}[(\beta U)^2]$,
we have
\begin{equation}\begin{aligned}
\Delta=1+\frac{1}{2}\beta U\,n(n-2)+\mathcal{O}\!\left[(\beta U)^2\right].
\label{eq:Delta_expand}
\end{aligned}\end{equation}
Thus the right side of Eq.~\eqref{eq:D_over_n_exact} can be simplified as
\begin{equation}\begin{aligned}
\frac{2-n}{n-1+\Delta}
=\frac{2-n}{n} +\frac{(2-n)^2}{2n}\,\beta U +\mathcal{O}\!\left[(\beta U)^2\right].
\label{eq:ratio_expand}
\end{aligned}\end{equation}
By substituting Eq.~\eqref{eq:ratio_expand} into Eq.~\eqref{eq:D_over_n_exact}, we obtain
\begin{equation}\begin{aligned}
\frac{D}{n}
&= \frac{1}{2}\Big[ \frac{2}{n} +\frac{(2-n)^2}{2n}\,\beta U \Big]^{-1} +\mathcal{O}[(\beta U)^2] \\
&= \frac{n}{4}\Big[1+\frac{(2-n)^2}{4}\beta U \Big]^{-1} +\mathcal{O}[(\beta U)^2] \\
&= \frac{n}{4} -\frac{n(2-n)^2}{16}\,\beta U +\mathcal{O}[(\beta U)^2],
\end{aligned}\end{equation}
in which the last equality is Eq.~(\ref{eq:AtomicD}). Besides, we also briefly discuss the fermion density at fixed $\mu$ under the atomic limit. For this case, both $\beta U$ and $\beta\mu$ are small numbers, and the expression of $n$ in Eq.~\eqref{eq:atomic_n_D_appendix} can be expanded as
\begin{equation}\begin{aligned}
n = 1-\frac{\beta U}{4}-\frac{\beta\mu}{2} +\mathcal{O}(\beta^2).
\label{eq:n_highT_fixedmu}
\end{aligned}\end{equation}
Since $\beta\mu = (\mu/U)\beta U$, the above result can be rewritten as
\begin{equation}
n = 1 - \frac{1+2\mu/U}{4}\,\beta U + \mathcal{O}(\beta^2).
\label{eq:n_highT_fixedmu_alt}
\end{equation}
This means that, at fixed (and finite) $\mu$, the filling $n$ generally approaches $n=1$ in the $T\to\infty$ limit, with the leading correction determined by $\mu/U$ and $\beta U$.

For the model~(\ref{eq:Hamiltonian}), $\mu=0$ corresponds to the half-filling case with $n=1$. In Sec.~\ref{Sec:Testin3D}, we have established the relation $\eyps(+\mu)=\eyps(-\mu)$ (with $\mu>0$) based on Eq.~(\ref{eq:ParSParMu}) and qualitative arguments. Here in the following, we prove it directly via the entropy evaluation formula in Eq.~(\ref{eq:changem}). First of all, the model Hamiltonian in Eq.~(\ref{eq:Hamiltonian}) with $+\mu$ and $-\mu$, denoted as $\hat{H}(U,\pm\mu)$, are connected via the particle-hole (PH) transformations [$c_{\mathbf{i}\sigma}^{+}\to(-1)^{\mathbf{i}}c_{\mathbf{i}\sigma}^{},c_{\mathbf{i}\sigma}^{}\to(-1)^{\mathbf{i}}c_{\mathbf{i}\sigma}^{+}$] as 
\begin{equation}\begin{aligned}
\hat{H}(U,+\mu) \overset{\text{PH}}{\Longrightarrow} \hat{H}(U,-\mu) + 2\mu N_s,
\end{aligned}\end{equation}
which indicates that the energy density $e_{\rm tot}$ should satisfy the relation $e_{\rm tot}(+\mu)=e_{\rm tot}(-\mu)+2\mu$. Similarly from the PH transformation, we have $n(-\mu)=2-n(+\mu)$, satisfying the $n(\mu=0)=1$ condition (as half filling). From Eq.~(\ref{eq:ParSParMu}), we compute $\eyps(+\mu)$ via
\begin{equation}\begin{aligned}
\label{eq:changemA1}
\eyps(+\mu) = \eyps(0) + \frac{1}{T}\Big[ e_{\mathrm{tot}}(+\mu) - e_{\mathrm{tot}}(0) -\int_{0}^{\mu} n(\mu^{\prime}) \dd{\mu^{\prime}} \Big].
\end{aligned}\end{equation}
For $\eyps(-\mu)$, we assume $\mu>0$ and accordingly reach
\begin{equation}\begin{aligned}
\label{eq:changemA2}
\eyps(-\mu) 
&= \eyps(0) + \frac{1}{T}\Big[ e_{\mathrm{tot}}(-\mu) - e_{\mathrm{tot}}(0) -\int_{0}^{-\mu} n(\mu^{\prime}) \dd{\mu^{\prime}} \Big] \\
&= \eyps(0) + \frac{1}{T}\Big[ e_{\mathrm{tot}}(-\mu) - e_{\mathrm{tot}}(0) + \int_{0}^{\mu} n(-\mu^{\prime}) \dd{\mu^{\prime}} \Big] \\
&= \eyps(0) + \frac{1}{T}\Big[ e_{\mathrm{tot}}(+\mu) - e_{\mathrm{tot}}(0) - \int_{0}^{\mu} n(\mu^{\prime}) \dd{\mu^{\prime}} \Big],
\end{aligned}\end{equation}
in which the relations $e_{\rm tot}(-\mu)=e_{\rm tot}(+\mu)-2\mu$ and $n(-\mu^{\prime})=2-n(+\mu^{\prime})$ are used. By comparing Eqs.~(\ref{eq:changemA1}) and~(\ref{eq:changemA2}), given that the same $\eyps(0)$ is applied, the equality $\eyps(+\mu)=\eyps(-\mu)$ is evident.

\section{Parameter grids and Monte Carlo sampling details}
\label{Sec:A3Simulation}

In this Appendix, we summarize the parameter grids and Monte Carlo sampling details used in our AFQMC simulations for different evaluation schemes presented in Sec.~\ref{Sec:Numeric}. For the temperature-integration schemes, the hybrid $T$-$\beta$ integrals are split at $T_0/t=0.5$ (and $\beta_0t=2$). In the low-temperature part, the $T/t$ values range from the lowest simulated temperature up to $T_0/t=0.5$ at intervals of $0.05$. In the high-temperature part, the grid is defined in terms of the inverse temperature, with intervals of $0.2$ in $\beta t$ for $1\leq\beta t <2$ and $0.1$ for $0\leq\beta t <1$. We have also verified that other choices of $T_0/t$ yield consistent entropy results within statistical uncertainties. For the varying-$U$ calculations, we adopt $U/t$ values at intervals of $0.5$. For the varying-$\mu$ calculations with repulsive interactions, the $\mu/t$ intervals are $0.2$, $0.12$, and $0.06$ for $U/t=6$, $4$, and $2$, respectively, whereas an alternative grid of $0.02$ in $\mu/t$ is applied for calculations with attractive interactions.

For a specific set of parameters (i.e., $U/t$, $\mu/t$, and $T/t$), all the Monte Carlo samples are organized into 23 statistically independent bins, with each bin constructed from 64 independently seeded Markov chains. Mostly, each Markov chain contains $\sim 1000$ measurement sweeps, giving a total of $\sim10^6$ measurement sweeps. For instance, at $U/t=6$, $T/t=0.3$, and $n=0.875$, each chain includes 2000 measurement sweeps, which corresponds to a total of $\sim 3\times10^6$ ordinary sweeps. One ordinary sweep consists of two sequential scans through all imaginary-time slices, first from $\tau=\beta$ to $\tau=0$ and then from $\tau=0$ to $\tau=\beta$. One $\tau$-line global update is performed at $\tau=0$ of each ordinary sweep for a randomly chosen lattice site.

\bibliography{EntropyRef}
\end{document}